\documentclass[aps,prd,twocolumn,showpacs,floatfix,preprintnumbers,amsmath,amssymb,nofootinbib, superscriptaddress]{revtex4-1}

\input epsf
\usepackage{graphicx}
\usepackage{color}

\newcommand{\beq}{\begin{equation}}
\newcommand{\eeq}{\end{equation}}
\newcommand{\barr}{\begin{eqnarray}}
\newcommand{\earr}{\end{eqnarray}}

\newcommand{\rme}{\textrm{e}}
\newcommand{\rmH}{\textrm{H}}
\newcommand{\Ly}{\textrm{Ly}}
\newcommand{\pabn}{p_{\textrm{ab}}^n}
\newcommand{\pscn}{p_{\textrm{sc}}^n}
\newcommand{\rmd}{\textrm{d}}
\newcommand{\N}{\mathcal{N}}
\newcommand{\nuc}{\nu_{\rm c}}

\newcommand{\apjl}{Astrophys. J. Lett.}
\newcommand{\aap}{Astron. Astrophys.}
\newcommand{\apjs}{Astrophys. J. Suppl. Ser.}

\newcommand{\aj}{Astron. J.}

\newcommand{\mnras}{Mon. Not. R. Astron. Soc.}

\newcommand{\memras}{Mem. R. Astr. Soc. }
\newcommand{\physrep}{Phys. Rep.}
\newcommand{\araa}{Annual Rev. Astron. Astrophys.}

\newcommand{\lsim}{\mathrel{\hbox{\rlap{\lower.55ex\hbox{$\sim$}} \kern-.3em \raise.4ex \hbox{$<$}}}}
\newcommand{\gsim}{\mathrel{\hbox{\rlap{\lower.55ex\hbox{$\sim$}} \kern-.3em \raise.4ex \hbox{$>$}}}}
\begin{document}
\title{Radiative transfer effects in primordial hydrogen recombination}

\author{Yacine Ali-Ha\"imoud}
\affiliation{California Institute of Technology, Mail Code 350-17, Pasadena, CA 91125}
\author{Daniel Grin}
\affiliation{California Institute of Technology, Mail Code 350-17, Pasadena, CA 91125}
\affiliation{Institute for Advanced Study, Einstein Drive, Princeton, NJ 08540, USA}
\author{Christopher M. Hirata} 
\affiliation{California Institute of Technology, Mail Code 350-17, Pasadena, CA 91125}

\date{\today}
\begin{abstract}
The calculation of a highly accurate cosmological recombination history has been the object of particular attention recently, as it constitutes the major theoretical uncertainty when predicting the angular power spectrum of Cosmic Microwave Background anisotropies. Lyman transitions, in particular the Lyman-$\alpha$ line, have long been recognized as one of the bottlenecks of recombination, due to their very low escape probabilities. The Sobolev approximation does not describe radiative transfer in the vicinity of Lyman lines to a sufficient degree of accuracy, and several corrections have already been computed in other works. In this paper, the impact of some previously ignored radiative transfer effects is calculated. First, the effect of Thomson scattering in the vicinity of the Lyman-$\alpha$ line is evaluated, using a full redistribution kernel incorporated into a radiative transfer code. The effect of feedback of distortions generated by the optically thick deuterium Lyman-$\alpha$ line blueward of the hydrogen line is investigated with an analytic approximation. It is shown that both effects are negligible during cosmological hydrogen recombination. Secondly, the importance of high-lying, non overlapping Lyman transitions is assessed. It is shown that escape from lines above Ly$\gamma$ and frequency diffusion in Ly$\beta$ and higher lines can be neglected without loss of accuracy. Thirdly, a formalism generalizing the Sobolev approximation is developed to account for the overlap of the high-lying Lyman lines, which is shown to lead to negligible changes to the recombination history.  Finally, the possibility of a cosmological hydrogen recombination maser is investigated. It is shown that there is no such maser in the purely radiative treatment presented here.
\end{abstract}

\maketitle

\section{Introduction}
The first measurements of the Cosmic Microwave Background (CMB) spectrum \cite{Mather90} and temperature anisotropies \cite{Smoot92}
changed cosmology from a qualitative to a robust and predictive science. Since then our picture of the Universe has become more and more accurate. Observations of high-redshift type Ia supernovae \cite{SN_Riess, SN_Perlmutter} have made it clear that nearly three fourths of the energy budget of our Universe is a non-clustering ``dark energy'' fluid with a negative pressure. In the last decade, the measurements of the temperature and polarization anisotropies in the CMB by the Wilkinson Microwave Anisotropy Probe (WMAP) \cite{WMAP7} have confirmed this picture and propelled cosmology into the era of high precision. Combined with other CMB measurements (e.g. BOOMERANG \cite{Boomerang}, CBI \cite{CBI}, ACBAR \cite{ACBAR}, QUaD \cite{QUAD}) and large-scale structure surveys (2dF \cite{2dF}, SDSS \cite{SDSS}), WMAP results have firmly established the $\Lambda$CDM model as the standard picture of our Universe.

What has also emerged from this high-precision data is our ignorance of the large majority of the constituents of the Universe. Only $\sim 4\%$ of our Universe is in the form of known matter (most of which is not luminous), the rest is in the form of an unknown clustering ``dark matter'' ($\sim 23\%$) or the even more disconcerting ``dark energy'' ($\sim 73 \%$). In addition, it is now widely believed that the Universe underwent an \emph{inflationary} phase early on that sourced the nearly scale-invariant primordial density perturbations, which led to the large-scale structure we observe today. Inflation requires non-standard physics, and at present there is no consensus on the mechanism that made the Universe inflate, and only few constraints on the numerous inflationary models are available from observations. 

The \emph{Planck} satellite, launched in May 2009, will measure the power spectrum of temperature anisotropies in the CMB, $C_{\ell}^{\rm T T}$, with a sub-percent accuracy, up to multipole moments $\ell \sim 2500$ \cite{Planck}. It will also measure the power spectrum of E-mode polarization anisotropies up to $\ell \sim 1500$. With this unprecedented ultra-high-precision data, cosmologists will be in a position to infer cosmological parameters accurate to the sub-percent level. The high resolution of \emph{Planck} observations will provide a lever arm to precisely measure the spectral index of scalar density perturbations $n_{\rm S}$ and their running $\alpha_{\rm S}$, therefore usefully constraining models of inflation. The polarization data will help break degeneracies of cosmological parameters with the optical depth to the surface of last scattering $\tau$, giving us a better handle on the epoch of reionization. This wealth of upcoming high-precision data from \emph{Planck}, as well as that from ongoing experiments (ACT \cite{ACT}, SPT \cite{SPT}) or possible future space-based polarization missions (CMBPol \cite{CMBPol}), can be fully exploited only if our theoretical predictions of CMB anisotropies are at least as accurate as the data.

The physics of CMB anisotropy generation is now well understood, and public Boltzmann codes are available (\textsc{CMBFast} \cite{CMBFAST}, CAMB \cite{CAMB}, \textsc{CMBEasy} \cite{CMBEASY}), which evolve the linear equations of matter and radiation perturbations and output highly accurate CMB temperature and polarization angular power spectra, for a given ionization history \cite{Boltzmann}. The dominant source of systematic uncertainty in the predicted $C_{\ell}$s is the recombination history \cite{Hu1995}. Not only the peak and width of the visibility function are important, but the precise shape of its tails is also critical at the sub-percent level of accuracy, in particular for the Silk damping tail \cite{Hu_Silk} of the anisotropy power spectrum. This has motivated Seager et al. \cite{Recfast_long, Recfast_short} to revise the seminal work of Peebles \cite{Peebles} and Zeldovich et al. \cite{Zeldovich_et_al} and extend their three-level atom model to a multi-level atom (MLA) calculation. Their recombination code, \textsc{RecFast}, is accurate to the percent level, and is a part of the Boltzmann codes routinely used for current day CMB data analysis. While sufficiently accurate for WMAP data, \textsc{RecFast} does not satisfy the level of accuracy required by \emph{Planck} \cite{Lewis06, Chluba10_uncertainties}. 

In the last few years, significant work has been devoted to further understanding the rich physics of cosmological recombination. Helium recombination is now understood to a sufficient level of accuracy for \emph{Planck} \cite{Hirata_SwitzerI, Hirata_SwitzerII, Hirata_SwitzerIII, Kholupenko_He, Wong_Scott07, Rubino08_He, Kholupenko08_He, CS10_He}. Cosmological hydrogen recombination demands a much higher level of accuracy than helium \cite{Wong08}, and its precise description is still the subject of ongoing efforts. Previous work on hydrogen recombination can be cast into two categories.

On the one hand, accurate recombination histories need to account for as large a number as possible of excited states of hydrogen. This is particularly important at late times, $z \lesssim 800-900$, when the free electron abundance becomes very low and the slow recombinations to the excited states become the ``bottleneck'' of the recombination process. In these conditions, it is important to precisely account for all possible recombination pathways by including a large number of excited states in MLA calculations. Since the recombination rates strongly depend on the angular momentum quantum number $l$, an accurate code must resolve the angular momentum substates \cite{RMCS06, CRMS07} and lift the statistical equilibrium assumption previously made. The standard MLA approach requires solving for the population of all the excited states accounted for, which is computationally expensive and has limited recent high-$n$ computations \cite{Grin_Hirata, Chluba_Vasil} to only a few points in parameter space. Recently, a new effective MLA (EMLA) method has been introduced \cite{EMLA}, which makes it possible to account for a very large number of excited states, while preserving the computational efficiency of a simple few-level atom. The EMLA approach consists in factoring the effect of the ``interior'' excited states (states which are not connected to the ground state) into effective recombination and photoionization coefficients and bound bound transition rates for the small number of ``interface'' states radiatively connected to the ground state, i.e. $2s, 2p$ and the low-lying $p$ states.

The second category of studies has concentrated on the transitions from the ``interface'' states to the ground state, of particular importance at early times when the overall recombination rate is controlled by the so called ``$n$=2 bottleneck''. Hydrogen atoms in the $n$=2 shell can reach the ground state either through emissions of single photons from the $2p$ state, which slowly escape the optically thick Lyman-$\alpha$ transition through cosmological redshifting, or from the $2s$ state, through forbidden two-photon decays. Previous studies have examined a series of effects that may affect these transition rates, including but not limited to stimulated $2s\rightarrow 1s$ two-photon decays and non-thermal $1s\rightarrow 2s$ two-photon absorptions \cite{CS06, Kholupenko06, Hirata_2photon}, feedback between neighboring lines of the optically thick Lyman series \cite{CS07, Kholupenko_Deuterium}, time-dependent effects in the Lyman-$\alpha$ line \cite{CS09a, Hirata_2photon}, two-photon decays from higher excited states \cite{Dubrovich_Grachev05, CS_2photon, Hirata_2photon, CS09b} and diffusion in the Lyman-$\alpha$ line \cite{Dubrovich_Grachev08, CS09c, Hirata_Forbes}.  

The purpose of this work is mainly to assess the importance of a few radiative transfer effects in the Lyman lines, that have not been investigated yet, or not in sufficient detail. In Section \ref{section:rad trans}, we review the theory of radiative transfer in the vicinity of a resonant line in an expanding Universe. We then turn to the Lyman-$\alpha$ line, for which we study the effect of Thomson scattering (Section \ref{section:Thomson}) and the interaction with the Ly$\alpha$ line of deuterium (Section \ref{section:deuterium}). In Section \ref{section:resonant scat}, we quantify the importance of the high-lying, non-overlapping Lyman transitions. Section \ref{section:overlap} is devoted to the overlapping high-lying Lyman lines. We also explore the possibility of a cosmological recombination maser in Section \ref{minimestepawayfromthelaser}. We summarize and discuss our results in Section \ref{section:discussion}. In this paper we will not repeat the standard MLA nor the EMLA formalism, and refer the reader to Refs.~\cite{Recfast_long, Hirata_2photon, Grin_Hirata} and \cite{EMLA}, respectively, for a detailed description.

\section{Radiative transfer in the Lyman lines} \label{section:rad trans}
\subsection{Basic notation}

\begin{table*}
\caption{
        \label{tab:notation}
        Notation used in this paper. Units of ``1'' means the quantity is dimensionless.
        }
\begin{tabular}{cccc}
\hline\hline
Symbol                   &Units     &Description     &Equation         \\
\hline
$A_{np,1s}$ & s$^{-1}$ & Einstein A-coefficient for the spontaneous decay $np\rightarrow 1s$ & Eq.~(\ref{eq:Anp1s}) \\
$a_n$  & 1 & dimensionless Voigt parameter for the Ly-$n$ transition & $a_n = \Gamma_{np}/(4 \pi \nu_n \Delta_{\rm H})$ \\
$f_{\nu}$ &1 & photon occupation number at frequency $\nu$ &  \\
$f_{\rm eq}^n$ &1& equilibrium value of $f_{\nu}$ near Ly-$n$  & $f_{\rm eq}^n = x_{np}/(3 x_{1s})$ \\
$f_{(\rm em)}^n$ & 1 & equilibrium value of $f_{\nu}$ near Ly-$n$, with true absorption and emision only & Eq.~(\ref{eq:fn0})\\
$H$ & s$^{-1}$ & Hubble expansion rate &\\
$N_{\rm H}$   &cm$^{-3}$     &total number density of hydrogen nuclei    &       \\
$N_{\rm X}$   &cm$^{-3}$     &number density of species X    &       \\
$\N_{\nu}$ &Hz$^{-1}$& number of photons per hydrogen atom per unit frequency & $\N_{\nu} = \frac{8 \pi \nu^2}{c^3 N_{\rm H}} f_{\nu}$ \\
$p_{\rm sc}^n$ & 1& fraction of photon absorptions in Ly-$n$ resulting in a scattering & $p_{\rm sc}^n = A_{np,1s}/\Gamma_{np}$ \\
$p_{\rm ab}^n$ &1& fraction of true photon absorptions in Ly-$n$ &$p_{\rm ab}^n = 1 - p_{\rm sc}^n$ \\
$R_{n'l\rightarrow np}$ & s$^{-1}$ & radiative transition rate $n'l \rightarrow np$ per hydrogen atom in the $[n',l]$ state&  \\
$R_n(\nu,\nu')$&Hz$^{-2}$ & resonant scattering redistribution kernel in Ly-$n$ & Eq.(\ref{eq:res.scat.kernel})\\
$R_{\rm T}(\nu \rightarrow \nu')$ & Hz$^{-1}$& Thomson scattering kernel & Eq.~(\ref{eq:rad_trans_T})\\
$\mathcal{R}(\delta)$ &1& dimensionless Thomson scattering kernel & Eq.~(\ref{eq:Thomson R unitless})\\
$\mathcal{S}_n$ &Hz&  width over which resonant scattering is effective in Ly-$n$ & Eq.~(\ref{eq:Sn})\\
$\mathcal{S}$ &Hz&  width over which resonant scattering is effective in Ly$\alpha$ & $\mathcal{S} = \mathcal{S}_2$\\
$S$       &1& dimensionless width over which resonant scattering is effective in Ly$\alpha$ & $S = \left(h \mathcal{S}/(k T_{\rm r})\right)^3$ \\
$T_{\rm m}, T_{\rm r}$ & K& matter and radiation temperatures & \\
$\mathcal{W}_n$ &Hz&width over which the Ly-$n$ wings are optically thick for true absorption& Eq.~(\ref{eq:Wn})\\
$\mathcal{W}$ &Hz&width over which the Ly$\alpha$ wings are optically thick for true absorption& $\mathcal{W} = \mathcal{W}_2$\\
$W$             &1& dimensionless width over which Ly$\alpha$ is optically thick for true absorption & $W = h \mathcal{W}/(k T_{\rm r})$\\
$x_{\rm D}$   & 1                  &abundance of deuterium relative to total hydrogen nuclei &  $x_{\rm D} = N_{\rm D}/N_{\rm H}$\\
$x_{nl}$       & 1                  &fraction of hydrogen in the state $[n,l]$ &       $x_{nl} = N_{\textrm{H}(n,l)}/N_{\rm H}$ \\
$x_e$         & 1                  & abundance of free electrons relative to total hydrogen nuclei & $x_e = N_e/N_{\rm H}$\\
$x$ &1& detuning from line center (when considering a single line),  & $ x= (\nu - \nu_n)/(\nu_n \Delta_{\rm H})$\\
      &  & or from $\nu_c$ (when considering line overlap), in Doppler units & $ x= (\nu - \nu_c)/(\nu_c \Delta_{\rm H})$\\
$x_n$ &1& detuning between Ly-$n$ and Ly-continuum frequencies in Doppler widths & $ x_n= (\nu_n - \nu_c)/(\nu_c \Delta_{\rm H})$\\
$y$ &1& same as $x$ OR detuning from the Ly$\alpha$ frequency in units of $kT_{\rm r}/h$ & $ y = h(\nu - \nu_{\Ly \alpha})/(kT_{\rm r})$\\
$y_{\rm D}$ &1& detuning of the H and D Ly$\alpha$ frequencies in units of $kT_{\rm r}/h$ & $ y_{\rm D} = h(\nu_{\rm D} - \nu_{\rm H})/(kT_{\rm r})$\\
$\Gamma_{np}$& s$^{-1}$ & width (or inverse lifetime) of the $np$ state & \\
$\Delta_{\rm X}$ &1& dimensionless Doppler width for a scattering species of mass $m_{\rm X}$ & $\Delta_{\rm X} = \sqrt{2 k T_{\rm m}/(m_{\rm X} c^2)}$ \\
$\Delta \nu_e$ &Hz & rms frequency shift during an electron scattering event near Ly$\alpha$& $\Delta \nu_e = \nu_{\textrm{Ly}\alpha} \Delta_e$\\ 
$\Delta \Psi_{\rm D}$ &1 & relative amplitude of the distortion due to deuterium at D Ly$\alpha$ & Eq.~(\ref{eq:DPsiD.def})\\ 
$\eta_e$ &Hz$^{-1}$ & differential optical depth for Thomson scattering near Lyman-$\alpha$ & Eq.~(\ref{eq:eta_e})\\
$\nu_{\Ly \alpha}$ ($\nu_{\rm H}, \nu_{\rm D}$) & Hz & Lyman-$\alpha$ resonant frequency (in hydrogen or deuterium specifically)& \\
$\nu_c$ & Hz& Lyman-limit frequency & $\nu_c = \frac43 \nu_{\Ly \alpha}$\\
$\nu_n$ &Hz & Lyman-$n$ resonant frequency & $\nu_n =(1 - n^{-2})\nu_c$\\ 
$\sigma_{\rm T}$ &cm$^2$& Thomson cross section & \\
$\sigma_0$ &cm$^2$ & photoionization cross section from the ground state, at threshold & Eq.~(\ref{eq:sigma0})\\
$\tau_n$     &1& Sobolev optical depth in the Lyman-$n$ resonance line & Eq.~(\ref{eq:tau_n def}) \\
$\tau_{\Ly \alpha}$     &1& Sobolev optical depth in the Lyman-$\alpha$ line & $\tau_{\Ly \alpha} = \tau_2$\\
$\tau_c$ & 1  & Lyman continuum optical depth per Doppler width near threshold & Eq.~(\ref{eq:tau_c def})\\
$\varphi_n(\nu)$& Hz$^{-1}$ & line profile for the Ly-$n$ transition & \\
$\phi_{\rm V}(x;a)$ & 1 &dimensionless Voigt profile, with Voigt parameter $a$ & Eq.~(\ref{eq:phi Voigt})\\
$\phi_n(x)$ &1& dimensionless Doppler profile centered at the Ly-$n$ frequency & $\phi_n(x) = \pi^{-1/2}\rme^{-(x - x_n)^2}$ \\
$\phi_c(x)$ &1& non-normalized dimensionless profile for continnum absorption near threshold & Eq.~(\ref{eq:phi_c}) \\
$\Psi^{(\rm H)}(y)$ &1& dimensionless photon occupation number near Ly$\alpha$ whith hydrogen only& Eqs.~(\ref{eq:fnu.change.var}), (\ref{eq:Psi_H}) \\
$\Psi^{(\rm H+D)}(y)$ &1& dimensionless photon occupation number near Ly$\alpha$ whith deuterium& Eqs.~(\ref{eq:fnu.change.var}), (\ref{eq:Psi_D}) \\
$\Psi_{\rm D}$ &1& dimensionless equilibrium photon occupation number near D Ly$\alpha$ & Eq.~(\ref{eq:PsiD.def})\\
$\psi(y)$ &1& rescaled difference in photon occupation numbers with or without deuterium & Eqs.~(\ref{eq:psi def}), (\ref{eq:psi}) \\

\hline\hline
\end{tabular}
\end{table*}

In this section, we present the basic quantities and notation used throughout this paper. We summarize our notation in Table~\ref{tab:notation}.

The photon occupation number at frequency $\nu$ is denoted $f_{\nu}$. In the case of a black body spectrum with temperature $T$, $f_{\nu}= \left(\rme^{h \nu/k T} - 1\right)^{-1}$. We will also make use of the number of photons per hydrogen atom per unit frequency, 
\beq
\N_{\nu} \equiv \frac{8 \pi \nu^2}{c^3 N_{\rm H}} f_{\nu},
\eeq 
where  $N_{\rm H}$ is the number density of hydrogen atoms. The population of a species X relative to the total abundance of hydrogen is denoted $x_{\rm X} \equiv N_{\rm X}/N_{\rmH}$. The fractional abundance of hydrogen in the state $[n,l]$ is denoted $x_{nl}$. For the low $l$ states, we use the spectroscopic notation $s,p,d,...$, so the ground state is denoted $1s$ and the $[n, l=1]$ states are denoted $np$.

This work will be concerned primarily with the $np\rightarrow 1s$ transitions, which will be referred to as the Lyman-$n$ (or Ly-$n$) transitions. The Ly-$2$ transition therefore designates, in that convention, the Lyman-$\alpha$ (Ly$\alpha$) transition. We denote the resonant Ly-$n$ transition frequency 
\beq
\nu_n \equiv \frac{4}{3}\left(1 - \frac{1}{n^2}\right) \nu_{\Ly \alpha} = \left(1 - \frac1{n^2}\right)\nu_c,
\label{eq:nu_n}
\eeq
where $\nu_{\Ly\alpha} \approx 2.47 \times 10^{15}$ Hz is the Lyman $\alpha$ frequency, and $\nu_c = \frac43 \nu_{\Ly \alpha}$ is the Lyman-limit frequency. The spontaneous emission rate (Einstein A-coefficient) in the Ly-$n$ transition is \cite{Bethe}:
\barr
A_{np,1s} &=& \frac{2^{13}\pi^3}{3^2 n^3} \frac{\left(1 -\frac1n\right)^{2n-5}}{\left(1 +\frac1n\right)^{2n+5}} \frac{\nu_n^3}{c^2}  \alpha a_0^2 \label{eq:Anp1s}\\
&\underset{n \gg 1}{\sim} &\frac{2^{13}\pi^3}{3^2 \exp(4)} \frac1{n^3}\frac{\nu_c^3}{c^2}  \alpha a_0^2. \label{eq:Anp1s large n}
\earr
where $\alpha$ is the fine structure constant and $a_0$ is the Bohr radius. 

We define the ratio 
\beq
f_{\rm eq}^n \equiv \frac{x_{np}}{3 x_{1s}},
\eeq
which is the equilibrium value of the photon occupation number at the Ly-$n$ transition frequency.

The Sobolev optical depth for the hydrogen Ly-$n$ transition is \cite{dellAntonio-Rybicki93, Recfast_long}:
\beq
\tau_n = \frac{3 c^3 N_{\rmH} x_{1s}}{8 \pi H \nu_n^3} A_{np,1s}\left(1 - \frac{x_{np}}{3 x_{1s}}\right),
\label{eq:tau_n def}
\eeq
where $H(z)$ is the Hubble expansion rate. In all that follows, we will neglect stimulated emission in the Lyman lines, as the photon occupation number near Ly-$n$ is of order $f_{\rm eq}^n \ll 1$ (the largest value is for $n = 2$ and is less than $10^{-11}$ for $z < 1600$). In particular, we can neglect the last term\footnote{When dealing with the possibility of cosmological masers in Sec.~\ref{minimestepawayfromthelaser}, we will of course explicitly account for this term and use the appropriate expression for the Sobolev optical depth in a general (not necessarily Lyman) transition.} in the expression for the Sobolev optical depth Eq.~(\ref{eq:tau_n def}).

Finally, we will refer to the matter temperature as $T_{\rm m}$ and the radiation temperature as $T_{\rm r}$. In practice, the matter temperature is locked to the radiation temperature through Thomson scattering, and the relative difference between the two is below a percent until redshift $z \approx 500$ \cite{Recfast_long, Hirata_SwitzerI}. 

\subsection{Line processes}
Consider an excited hydrogen atom in the $np$ state; it has two mutually exclusive fates.

The first possibility is that it reaches another excited state $n's$ or $n'd$, with $n'\ne 1$, either through a spontaneous or stimulated decay if $n' < n$ or following the absorption of a CMB photon if $n' > n$. It can also be photoionized by a CMB photon. 

The second possibility is that the atom spontaneously decays to the ground state, emitting a Ly-$n$ photon. In principle, this decay can also be stimulated. However, even when accounting for non thermal distortions to the radiation field, the photon occupation number at Lyman frequencies is extremely small, and stimulated emission in the Lyman lines can be neglected. 

The probabilities of these two complementary fates are denoted $\pabn, \pscn$ respectively (the justification of the notation will become clearer in the next paragraph). Given the width (or inverse lifetime) $\Gamma_{np}$ of the $np$ state (the sum of the rates of all transitions depopulating this state), they are given by:
\beq
\pscn = \frac{A_{np, 1s}}{\Gamma_{np}} = 1 - \pabn.
\label{eq:pscat}
\eeq
If we now assume that the considered atom was initially in the ground state and reached the $np$ state after the absorption of a resonant Ly-$n$ photon, the two fates mentioned above can be described in a two-photon picture.

First, if the atom reaches another excited state $n' < n$, the overall reaction $\rmH(1s) + \gamma($Ly-$ n)\rightarrow \rmH(n'l) + \gamma'$ is a (possibly stimulated) Raman scattering event. If the atom absorbs a CMB photon and reaches a higher excited state $n'> n$ (or gets photoionized), the overall reaction $\rmH(1s) + \gamma($Ly-$ n) + \gamma' \rightarrow \rmH(n'l)$ (or $\rightarrow e^- + p^+$) is a two-photon absorption (or two-photon photoionization) event. In these cases we will refer to the absorption of the Ly-$n$ photon as a true absorption event, in the sense that the photon is destroyed in the process. The emission of a Ly-$n$ photon following the inverse reaction chain will be referred to as a true emission event.

Secondly, if the atom decays back to the ground state, the overall reaction $\rmH(1s) + \gamma($Ly-$n) \rightarrow \rmH(1s) + \gamma($Ly-$n)$ is a Rayleigh scattering event (which in what follows we will refer to as a scattering event for short). In that case the incoming and outgoing photons have the same frequency in the atom's rest frame. Their frequencies in the comoving frame (frame in which the CMB appears isotropic) are Doppler-shifted with respect to the atom's rest frame frequencies. Since the Doppler shift depends on the relative orientation of the photon propagation direction and the atom's velocity, the frequencies of the incoming and outgoing photons in the comoving frame are in general different. They are however statistically correlated, as will be described in Section \ref{section:scattering}.

\subsubsection{True absorption and emission}
The rate of true emission of resonant Ly-$n$ photons at frequency $\nu$, per H atom, per frequency interval, is given by \cite{Hirata_SwitzerI, Hirata_Forbes}:
\barr
\dot{\N}_{\nu}\big{|}_{\rm em} &=& \left( \sum_{(n'\ne 1), l} x_{n'l} R_{n'l \rightarrow np} + x_e x_p N_{\rm H} \alpha_{np} \right)\nonumber\\
&\times& \pscn ~ \varphi_n(\nu). \label{eq:dotNem} 
\earr
In the above equation, $R_{n'l \rightarrow np}$ is the radiative transition rate per hydrogen atom from the $n' l$ state to the $np$ state, including stimulated transitions, $\alpha_{np}$ is the direct recombination coefficient to the $np$ state, including stimulated recombinations, and $\varphi_n(\nu) $ is the line profile, which has the Voigt shape\footnote{The Voigt profile can be derived quantum-mechanically, in the two-photon picture, when one neglects the variation of multiplicative factors $\nu/\nu_n$ across the line and uses the resonance approximation. See e.g. Ref.~\cite{Hirata_Forbes} for fits to the correct profile in the case of the Lyman $\alpha$ line.}:
\beq
\varphi_n(\nu) = \frac{1}{\nu_{n}\Delta_{\rm H}}\phi_{\rm V}\left(\frac{\nu - \nu_n}{\nu_n\Delta_{\rm H}} ; a_n\right),
\eeq
where 
\beq
 \phi_{\rm V}(x; a) \equiv \frac{a}{\pi^{3/2}}\int_{-\infty}^{+\infty} \frac{\rme^{-t^2}}{a^2 + (x - t)^2} \rmd t, \label{eq:phi Voigt}
\eeq 
is the dimensionless Voigt profile, 
\beq
\Delta_{\rm H} \equiv \sqrt{\frac{2 k T_{\rm m}}{m_{\rm H} c^2}} \approx 2.35 \times 10^{-5} \left(\frac{1+z}{1100}\frac{T_{\rm m}}{T_{\rm r}}\right)^{-1/2}
\eeq
is the dimensionless Doppler width and 
\beq
a_n = \frac{\Gamma_{np}}{4 \pi \nu_n\Delta_{\rm H}} = \frac{1}{\pscn}\frac{A_{np,1s}}{4 \pi \nu_n \Delta_{\rm H}}
\eeq
is the dimensionless Voigt parameter of the line.

For small Voigt parameters $a \ll 1$, which is the case in all Lyman lines at the epoch of recombination, the Voigt profile has the well known asymptotic behaviors in the line center and in the damping wings:
\beq
\phi_{\rm V}(x;a) \approx \Bigg{\{}\begin{array}{ll}
\frac1{\sqrt \pi} \rme^{-x^2} &, x \lesssim x_a\\
\frac{a}{\pi x^2}  & , x \gtrsim x_a
\end{array} \label{eq:Voigt asymptotic}                     
\eeq
where the transition scale $x_a$ is the solution of \cite{Hummer_Voigt}:
\beq
x_a^2 \rme^{-x_a^2} = \frac{a}{\sqrt \pi}.
\eeq
In general, for $a \ll1 $, $x_a \sim 3$.

Following Ref.~\cite{Hirata_SwitzerI}, we define
\barr
f_{(\rm em)}^n \equiv \frac{\sum_{(n' \ne 1), l  } x_{n'l} R_{n'l \rightarrow np} + x_e x_p N_{\rm H}\alpha_{np} }{3 x_{1s}\Gamma_{np}\pabn},\label{eq:fn0}
\earr
so that the true emission rate per H atom per frequency interval can be rewritten as:
\beq
\dot{\N}_{\nu}\big{|}_{\rm em} =\pabn \ 3 x_{1s} A_{np,1s}\varphi_n(\nu)  f_{(\rm em)}^n.
\eeq
The rate of true absorption of resonant photons is simply the total rate of absorption times the true absorption probability. The absorption profile differs from the emission profile by a factor $\rme^{h(\nu - \nu_n)/(k T)}$, where $T = T_{\rm r}$ in the wings (because the low-energy photon of the two-photon process comes from a black-body distribution and the excited states of hydrogen are near Boltzmann equilibrium with each other at temperature $T_{\rm r}$ \cite{Hirata_2photon, Hirata_Forbes}), and $T = T_{\rm m}$ in the Doppler core, where atomic recoil tends to equilibrate the radiation field with the thermal velocity distribution of the atoms. The rate of true absorption of resonant Ly-$n$ photons at frequency $\nu$, per H atom, per frequency interval, is therefore given by:
\beq
\dot{\N}_{\nu}\big{|}_{\rm ab} = - \pabn 3 x_{1s} A_{np,1s} \varphi_n(\nu)  \rme^{\frac{h(\nu - \nu_n)}{k T}} f_{\nu} . 
\label{eq:dotxab}
\eeq
The net (uncompensated) rate of true emission of resonant Ly-$n$ photons at frequency $\nu$, per H atom, is therefore:
\barr
\dot{\N}_{\nu}\big{|}_{\rm em,ab} &=& \pabn \ 3 x_{1s} A_{np,1s} \varphi_n(\nu)\nonumber\\
 &\times&  \left[f_{(\rm em)}^n - \rme^{\frac{h(\nu - \nu_n)}{k T}} f_{\nu}\right].
\label{eq:dotNem,ab}
\earr
It will be useful in what follows to cast this expression into a different but equivalent form, which can be done with the following considerations.
 
The rate of change in the population of the $np$ state is:
\barr
\dot{x}_{np} &=& \sum_{n', l  } x_{n'l} R_{n'l \rightarrow np} + x_e x_p N_{\rm H} \alpha_{np} - x_{np}\Gamma_{np} \nonumber\\
&=&3 x_{1s}\Gamma_{np}\pabn f_{(\rm em)}^n + x_{1s}R_{1s\rightarrow np}- x_{np}\Gamma_{np}.\label{eq:total dotxnp}
\earr
The radiative rates (of order $\sim 10^8 $ s$^{-1}$) are many orders of magnitude larger than the overall recombination rate, which is of order the Hubble rate $H \sim10^{-13} $ s$^{-1}$. The population of the excited states can therefore be obtained to an excellent accuracy by using the steady state approximation and setting $\dot{x}_{np} = 0$ in the above equation. Setting the left hand side of Eq.~(\ref{eq:total dotxnp}) to zero, we can solve for $f_{(\rm em)}^n$:
\beq
f_{(\rm em)}^n = \frac{1}{3 x_{1s}\Gamma_{np}\pabn}\left[x_{np}\Gamma_{np} - x_{1s} R_{1s\rightarrow np} \right].
\eeq
The total (including both true absorptions and absorptions leading to a scattering)  $1s \rightarrow np$ (forward only) excitation rate per H atom is given by:
\beq
x_{1s} R_{1s \rightarrow np} = 3 x_{1s} A_{np,1s} \int \varphi_n(\nu)\rme^{\frac{h(\nu - \nu_n)}{k T}} f_{\nu} \rmd \nu.
\eeq
This finally gives us the following relation for $f_{(\rm em)}^n$ \cite{CS09a}:
\beq
\pabn f_{(\rm em)}^n = f_{\rm eq}^n - \pscn \int \varphi_n(\nu)\rme^{\frac{h(\nu - \nu_n)}{k T}} f_{\nu} \rmd \nu.
\label{eq2:fn0}
\eeq
We can now rewrite the net rate of true emission of resonant Ly-$n$ photons at frequency $\nu$, per H atom, in a form exactly equivalent to Eq.~(\ref{eq:dotNem,ab}):
\barr
&&\dot{\N}_{\nu}\big{|}_{\rm em,ab} = 3 x_{1s} A_{np,1s} \varphi_n(\nu) \Bigg{\{} f_{\rm eq}^n - \rme^{\frac{h(\nu - \nu_n)}{k T}} f_{\nu}
\nonumber\\
&& +\pscn  \left[ \rme^{\frac{h(\nu - \nu_n)}{k T}} f_{\nu} - \int \varphi_n(\nu')\rme^{\frac{h(\nu' - \nu_n)}{k T}} f_{\nu'} \rmd \nu' \right] \Bigg{\}}. \label{eq:dotNem,ab alternate}
\earr

\subsubsection{Coherent scattering} \label{section:scattering}
The rate at which resonant scattering removes photons from the line, at frequency $\nu$, (in photons per second per H atom per frequency interval) is the total absorption rate times the scattering probability:
\beq
\dot{\N}_{\nu}\big{|}_{\rm sc, -} = - \pscn 3 x_{1s} A_{np,1s} \varphi_n(\nu) \rme^{\frac{h(\nu - \nu_n)}{k T}} f_{\nu} . 
\label{eq:scat ab}
\eeq
The rate at which resonant scattering injects photons in the line, at frequency $\nu$, depends on the absorption at all other frequencies, since incoming and outgoing photon frequencies are correlated:
\barr
\dot{\N}_{\nu}\big{|}_{\rm sc, +} &=&  \pscn 3 x_{1s} A_{np,1s} \nonumber\\
&\times& \int p_n(\nu|\nu') \varphi_n(\nu') \rme^{\frac{h(\nu' - \nu_n)}{k T}} f_{\nu'}\rmd \nu',
\label{eq:scat em}
\earr
where $p_n(\nu|\nu')$ is the probability that the outgoing photon has frequency $\nu$ in the comoving frame given that the incoming photon had frequency $\nu'$. It accounts for the random thermal motions of the scattering atoms (at temperature $T_{\rm m}$) and depends on the angular probability distribution of a scattering event. It it is normalized:
\beq
{\rm for~all}\ \nu' ,   \ \int p_n(\nu|\nu') \rmd \nu = 1,
\eeq 
and to respect detailed balance, must satisfy
\beq
 p_n(\nu|\nu') \varphi_n(\nu')= p_n(\nu'|\nu)\varphi_n(\nu) .
\eeq
The scattering redistribution kernel 
\beq
R_n(\nu, \nu') \equiv p_n(\nu|\nu')\varphi_n(\nu') \rme^{\frac{h(\nu' - \nu_n)}{k T}} \label{eq:res.scat.kernel}
\eeq
is calculated in Ref.~\cite{Hummer62} (in which, however, atomic recoil during a scattering event is not accounted for). 

The most general form for the rate of change of the photon field through resonant scattering is given by Eqs.~(\ref{eq:scat ab}) and (\ref{eq:scat em}). However, in the case where the radiation field is smooth on the scale of a characteristic frequency shift in a scattering $\langle\Delta\nu^2\rangle^{1/2} = \nu_n \Delta_{\rm H} $, one can approximate the integral operator by a second-order differential operator. The rate of change of $\N_{\nu}$ due to scattering is then given by a Fokker-Planck equation \cite{Rybicki_DellAntonio, Hirata_WF, Rybicki06, Dubrovich_Grachev08, Hirata_Forbes, CS09c}, accounting for scattering as a diffusive process in frequency space, with a systematic shift (or drift) due to recoil and Doppler boosting:
\barr
\dot{\N}_{\nu}\big{|}_{\rm sc}^{\rm FP} &=& \pscn 3 x_{1s} A_{np,1s} \nonumber\\
&\times& \frac{\partial}{\partial \nu}\left\{\frac{\nu^2 \Delta_{\rm H}^2}2 \varphi_n(\nu)\left[\frac{\partial f_{\nu}}{\partial \nu}+ \frac{h }{k T_{\rm m}} f_{\nu}\right]\right\}. \label{eq:FP}
\earr
\subsubsection{The radiative transfer equation}\label{section:rad transfer}

In the vicinity of the Ly-$n$ line, the photon occupation number evolves under the influence of the resonant processes described above, as well as eventual non resonant processes that may act in the vicinity of the line. The time-dependent radiative transfer equation (the Boltzmann equation for the photon fluid) can be written in the general form:
\beq
\frac{\rmd f_{\nu}}{\rmd t} \equiv \frac{\partial f_{\nu}}{\partial t} - H \nu \frac{\partial f_{\nu}}{\partial \nu} = \dot{f}_{\nu}\big{|}_{\rm em,ab,sc} + \dot{f}_{\nu}\big{|}_{\rm nr}, \label{eq:rad trans}
\eeq
where $\rmd / \rmd t$ is the derivative along a photon trajectory, and $\dot{f}_{\nu}|_{\rm nr}$ groups all processes that are not resonant with the considered line (these could include absorption and emission from neighboring lines for example). The contribution of resonant processes is obtained with the conversion:
\barr
\dot{f}_{\nu}\big{|}_{\rm em,ab,sc} &=& \frac{c^3 N_{\rm H}}{8 \pi \nu^2} \dot{\N}_{\nu}\big{|}_{\rm em,ab,sc}.
\label{eq:conversion}
\earr
Neglecting the variation of multiplicative factors $\nu/\nu_n$ across the line, and using the definition of $\tau_n$, Eq.~(\ref{eq:tau_n def}), as well as Eq.~(\ref{eq:dotNem,ab alternate}), the most general expression for the resonant term can be written:
\barr
&&-\frac1{H\nu}\dot{f}_{\nu}\big{|}_{\rm em,ab,sc}  = \tau_n \varphi_n(\nu)\left[\rme^{\frac{h (\nu - \nu_n)}{k T}}f_{\nu}  - f_{\rm eq}^n \right] + \pscn \tau_n \nonumber\\
&&\times \int\left[\varphi_n(\nu) \varphi_n(\nu') \rme^{\frac{h(\nu' - \nu_n)}{k T}} - R_n(\nu, \nu')\right] f_{\nu'}\rmd \nu'. \label{eq:ab-em-sc alternate}
\earr
The radiative transfer equation, Eq.~(\ref{eq:rad trans}), with $\dot{f}_{\nu}|_{\rm em,ab,sc}$ given by Eq.~(\ref{eq:ab-em-sc alternate}), is therefore, in the general case, a time-dependent, partial integro-differential equation. As a result, it is computationally expensive to solve without further approximations.

If the radiation field varies on a frequency scale large compared to a Doppler width, then we can use the Fokker-Planck operator for the scattering term, and the above term can be approximated by:
\barr
&& -\frac1{H\nu}\dot{f}_{\nu}\big{|}_{\rm em,ab,sc}  \approx \pabn \tau_n \varphi_n(\nu)\left[\rme^{\frac{h (\nu - \nu_n)}{k T}}f_{\nu}  - f_{(\rm em)}^n \right] \nonumber\\
&&- \pscn \tau_n \frac{\partial}{\partial \nu}\left\{\frac{\nu^2 \Delta_{\rm H}^2}2 \varphi_n(\nu)\left[\frac{\partial f_{\nu}}{\partial \nu}+ \frac{h }{k T_{\rm m}} f_{\nu}\right] \right\}.
\label{eq:ab-em-sc-FP}
\earr
More insight can be gained by considering some characteristic scales of the problem. Using the asymptotic expansion for the wings of the Voigt profile, Eq.~(\ref{eq:Voigt asymptotic}), we obtain the total optical depth for true absorption in each damping wing: 
\beq
\tau_n^{\rm ab, wing} = \pabn \tau_n \frac{a_n}{\pi x_{a_n}}.
\eeq
If $\tau_{n}^{\rm ab, wing} \ll 1$, the damping wings are optically thin to true absorption, and one can use the Doppler core approximation to the Voigt profile in the radiative transfer equation. If $\tau_n^{\rm ab, wing} \gg 1$, then the wings are optically thick for true absorption up to a detuning from the line center:
\beq
\mathcal{W}_n \equiv \pabn \tau_n \frac{a_n}{\pi} \nu_n\Delta_{\rm H} \gg \nu_n \Delta_{\rm H}.
\label{eq:Wn}
\eeq
In that case, the radiation field is near its equilibrium value $f_{\nu} \approx f_{(\rm em)}^n \rme^{-h(\nu - \nu_n)/kT}$ within a detuning from the line center $|\nu - \nu_n| \lesssim \mathcal{W}_n$.
 
One can similarly define the optical depth for resonant scattering in each damping wing, $\tau_{n}^{\rm sc, wing}$. If $\tau_{n}^{\rm sc, wing} \gg 1$, we can see using dimensional analysis that resonant scattering, when described with a Fokker-Planck operator, is effective up to a characteristic detuning from the line center
\beq
\mathcal{S}_n \equiv \left( \pscn \tau_n \frac{a_n}{2\pi}\right)^{1/3} \nu_n \Delta_{\rm H}.
\label{eq:Sn}
\eeq
The physical meaning of this quantity can be understood as follows. A photon with initial frequency $\nu$ can diffuse (in frequency space) to the line center in a characteristic time
\beq
\Delta t_{\rm diff} \sim \frac{(\nu - \nu_n)^2}{(\nu_n \Delta_{\rm H})^2} \frac{1}{c x_{1s} N_{\rm H} \sigma_n(\nu)}, 
\eeq
where $\sigma_n(\nu)$ is the cross-section for resonant scattering:
\beq
\sigma_n(\nu) \equiv p_{\rm sc}^n \frac{3}{8 \pi} \frac{c^2}{\nu_n^2} A_{np,1s} \varphi_n(\nu).
\eeq
The time it takes for the photon to redshift from $\nu$ to $\nu_n$ (if $\nu > \nu_n$, or from $\nu_n$ to $\nu$ in the opposite case) is
\beq
\Delta t_{\rm redshift} = \frac{|\nu - \nu_n|}{\nu_n H}.
\eeq
Using the damping wings approximation for $\varphi_n(\nu)$ Eq.~(\ref{eq:Voigt asymptotic}), the definition of the Sobolev optical depth Eq.~(\ref{eq:tau_n def}), and Eq.~(\ref{eq:Sn}), we obtain:
\beq
\frac{\Delta t_{\rm diff}}{\Delta t_{\rm redshift}} \sim \left(\frac{|\nu - \nu_n|}{\mathcal{S}_n}\right)^3. \label{eq:S.meaning}
\eeq
Therefore the radiation field will reach the equilibrium spectrum $f_{\nu} \propto \rme^{- h \nu/ k T_{\rm m}}$ within a detuning from the line center $|\nu - \nu_n|\lesssim \mathcal{S}_n$, due to the very fast redistribution of photon frequencies through resonant scattering.

\begin{figure} 
\includegraphics[width = 85mm]{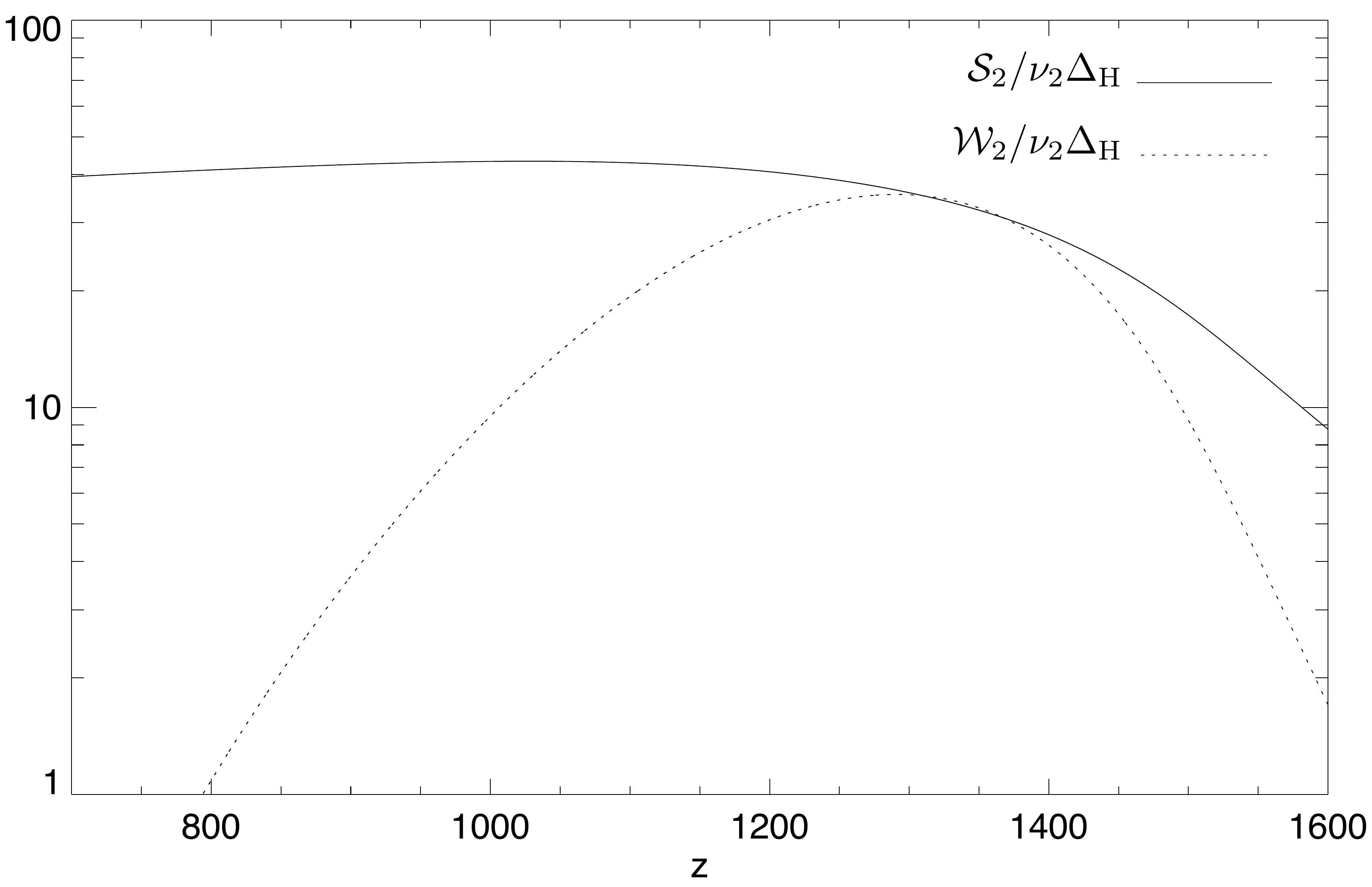}
\caption{Characteristic widths $\mathcal{W}, \mathcal{S}$ for the Lyman-$\alpha$ line, in units of Doppler widths, as a function of redshift for a standard recombination history.}
\label{fig:WS.Ly.alpha}
\end{figure}

As an illustration, we show in Fig.~\ref{fig:WS.Ly.alpha} the parameters $\mathcal{W}_2, \mathcal{S}_2$ for the Lyman-$\alpha$ line, extracted from the MLA code described in Ref.~\cite{Hirata_2photon}. We see that for $z \gtrsim 800$, $\mathcal{S}_2 \geq \mathcal{W}_2 > \nu_{\textrm{Ly} \alpha} \Delta_{\rm H}$ and at all relevant times $\mathcal{S}_2 \gg \nu_{\textrm{Ly} \alpha} \Delta_{\rm H}$. According to the above discussion, the radiation field in the vicinity of Ly$\alpha$ is therefore smooth on a frequency scale $\Delta \nu \sim \mathcal{S}_2$ around line center, and the use of the Fokker-Planck operator for resonant scattering is well justified.

\subsection{Net decay rate in the Lyman-$n$ line}
The exact shape of the radiation field in the vicinity of resonant lines is by itself of interest when predicting potentially observable spectral distortions to the black body spectrum of the CMB \cite{RMCS06}. In the context of cosmological recombination, the quantities of interest are the net (uncompensated) bound-bound and bound-free rates, which are required to evolve the atomic level populations and the free electron fraction in a MLA code. In particular, the net rate of $np \rightarrow 1s$ decays is given by:
\barr
\dot{x}_{np \rightarrow 1s} &=& \dot{x}_{1s}\big{|}_{np} = - \dot{x}_{np}\big{|}_{1s} \nonumber\\ 
&=& \frac{\rmd}{\rmd t}\int \N_{\nu}\rmd \nu =\int \dot{\N}_{\nu}\big{|}_{\rm em,ab} \rmd \nu.
\earr
Using Eq.~(\ref{eq:dotNem,ab alternate}), we see that this can be written:
\barr
\dot{x}_{np \rightarrow 1s} &=& 3 x_{1s} A_{np,1s} \nonumber \\
       &\times& \int\left[f_{\rm eq}^n - \rme^{\frac{h(\nu - \nu_n)}{k T}} f_{\nu}\right] \varphi_n(\nu) \rmd \nu,\label{eq:net decay simple}
\earr
where we used the fact that the term proportional to $\pscn$ in Eq.~(\ref{eq:dotNem,ab alternate}) integrates to zero.

As can be seen in Eq.~(\ref{eq:net decay simple}), the net decay rate in the Ly-$n$ transition depends on the radiation field. The latter in turns depends on the atomic level populations through $\tau_n, \pscn$ and $f_{\rm eq}^n$, as well on eventual non-resonant processes acting in the vicinity of the line, as can be seen from the radiative transfer equation Eqs.~(\ref{eq:rad trans}), (\ref{eq:ab-em-sc alternate}) or (\ref{eq:ab-em-sc-FP}) .

\subsection{The Sobolev approximation}
Equations (\ref{eq:rad trans}) and (\ref{eq:ab-em-sc alternate}) give the most general form\footnote{We neglected the variation of phase-space factors $\nu/\nu_n$ across the line, as well as stimulated emission and stimulated scatterings.} of the radiative transfer equation, in a homogeneous expanding Universe. It has no analytic solution, because of its complexity and since it requires the values of the level populations, which in turn depend on the radiation field. It can however be simplified and decoupled under some approximations, for which an analytic solution exists. The Sobolev approximation \cite{dellAntonio-Rybicki93, Recfast_long} relies on the following assumptions: \\
(i) \emph{No non-resonant processes} act in the vicinity of the line ($\dot{f}_{\nu}|_{\rm nr} = 0$ in Eq.~(\ref{eq:rad trans})).\\
(ii) \emph{Steady-state} : the time it takes a photon to redshift across the line is of order $w/H\nu_n$, where $w$ is the characteristic line width. If the line is very narrow, $w \ll \nu_n$, then this time is much smaller than the Hubble time, therefore physical quantities, such as $n_{\rm H}, T_{\rm r}, T_{\rm m}, H, x_e, x_{nl}$ vary very little during the time it takes a photon to redshift through the line. As a consequence, one can neglect the time dependence in Eq.~(\ref{eq:rad trans}).\\
(iii) \emph{Equal absorption and emission profile}. This assumption also derives from the assumption of an infinitesimally thin line, in which case one can take the exponential factors to be unity in Eq.~(\ref{eq:dotxab}).\\
(iv) \emph{Complete redistribution} of emitted photons. Mathematically, this means that $p_n(\nu|\nu')$ is independent of $\nu'$ in Eq.~(\ref{eq:scat em}). Because of assumption (iii), it is also assumed that $p_n(\nu|\nu') = \varphi_n(\nu)$, i.e. the scattered photons are completely redistributed over the line profile. This implies that $R_{n}(\nu, \nu') = \varphi_n(\nu) \varphi_n(\nu')$, and therefore the integral in Eq.~(\ref{eq:ab-em-sc alternate}) vanishes (taking the exponential to be unity).

The radiative transfer equation, under these assumptions, becomes the much simpler ordinary differential equation:
\beq
\frac{\rmd f_{\nu}}{\rmd \nu} = \tau_n \varphi_n(\nu)\left(f_{\nu} - f_{\rm eq}^n\right).
\label{eq:rad transfer sobolev}
\eeq
It has the analytic solution
\beq
f_{\nu} = f_{\rm eq}^n + (f_+^n - f_{\rm eq}^n)\exp\left[-\tau_n \int_{\nu}^{+\infty}\varphi_n(\nu') \rmd \nu' \right],
\eeq
where $f_+^n $ is the photon occupation number at the blue side of the line. The photon occupation number at the red side of the line is therefore
\beq
f_-^n = f_{\rm eq}^n + (f_+^n - f_{\rm eq}^n) \rme^{-\tau_n}.
\label{eq:f_-}
\eeq
Integrating Eq.~(\ref{eq:rad transfer sobolev}) from $-\infty$ to $+\infty$ gives
\beq
f_+^n - f_-^n = \tau_n \int\left[f_{\nu} - f_{\rm eq}^n\right]\varphi_n(\nu) \rmd \nu.\label{sobfinal}
\eeq
This, combined with Eqs.~(\ref{eq:net decay simple}) and (\ref{eq:f_-}) finally gives us the standard Sobolev expression for the net decay rate in the line:
\beq
\dot{x}_{np \rightarrow 1s} = 3 x_{1s} A_{np,1s} P_{\textrm{S},n} \left[f_{\rm eq}^n - f_+^n\right],
\label{eq:Sobolev rate}
\eeq
where
\beq
P_{\textrm{S},n} \equiv \frac{1 - \rme^{-\tau_n}}{\tau_n} 
\eeq 
is the Sobolev escape probability. In the case of Lyman transitions, $\tau_n \gg 1$, and $P_{\textrm{S},n} \approx 1/\tau_n$. The net decay rate becomes, using Eq.~(\ref{eq:tau_n def}):
\beq
\dot{x}_{np \rightarrow 1s} = \frac{8 \pi H \nu_n^3}{c^3 N_{\rm H}} \left[f_{\rm eq}^n - f_+^n\right],
\label{eq:Sobolev rate bis}
\eeq
which is simply the rate at which distortion photons redshift across the line.

The Sobolev approximation provides relatively accurate net decay rates, despite the multiple assumptions that it relies on, and recombination histories currently used for CMB analysis \cite{Recfast_long, Recfast_short}, which use this approximation, are accurate at the percent level. The level of precision required by upcoming CMB experiments has motivated recent work to obtain more accurate solutions to the radiative transfer equation and net bound-bound rates in the optically thick Lyman lines. In this work we investigate previously ignored radiative transfer effects, and quantify as much as possible the errors made by the inevitable approximations that still need to be made.

\section{The Lyman alpha line}
The Lyman-$\alpha$ transition is one of the bottlenecks of hydrogen recombination. Electrons recombine to the excited states of hydrogen, from which they eventually cascade down to the $n=2$ state. They can then reach the ground state either by a two-photon decay from the $2s$ state, with rate $\Lambda_{2s1s} \approx 8.22$ s$^{-1}$, or from the $2p$ state, by redshifting out of the Lyman-$\alpha$ resonance, with rate $A_{2p1s} P_{\rm esc}$, where $P_{\rm esc}$ is the escape probability. Due to its substantial impact on the recombination history (see for example Fig.~11 in Ref.~\cite{CS09a}), the net decay rate, or equivalently the escape probability in Ly$\alpha$ has been studied extensively, including time-dependent effects \cite{Rybicki_DellAntonio, CS09a}, two-photon processes \cite{Hirata_2photon, CS09b}, resonant scattering \cite{Krolik1989, Krolik1990, Dubrovich_Grachev08, CS09c, Hirata_Forbes}, and Thomson scattering \cite{CS09c}. In this section we consider two additional effects: the non-local aspect of Thomson scattering (Section \ref{section:Thomson}), and a quantitative estimate of the effect of deuterium on hydrogen recombination (Section \ref{section:deuterium}). 

To simplify the notation we drop the subscripts and superscripts ``2'' in this section, and all the quantities previously defined implicitly refer to Ly$\alpha$.

\subsection{Thomson scattering in Lyman-$\alpha$}\label{section:Thomson}
Thomson scattering in the vicinity of resonant lines was investigated in the context of helium recombination \cite{Hirata_SwitzerIII} with a Monte Carlo method and found to lead to negligible changes to the recombination history. Its effect on the hydrogen Ly$\alpha$ line was investigated recently \cite{CS09c} and found to lead to negligible corrections to the escape probability and the recombination history. In Ref.~\cite{CS09c} however, electron scattering was described with the Kompaneets equation, which is not valid in the context of cosmological hydrogen recombination, as we argue below. Here we provide a more rigorous treatment of Thomson scattering, using the full redistribution kernel, which we incorporate in the Lyman-$\alpha$ radiative transfer code described in Ref.~\cite{Hirata_Forbes}.

The strength of Thomson scattering is characterized by its differential optical depth, flat in frequency in the non relativistic limit \cite{Hirata_SwitzerIII}:
\beq
\eta_e \equiv \frac{N_{\rm H} x_{\rme} \sigma_{\rm T} c}{H \nu_{\Ly \alpha}},
\label{eq:eta_e}
\eeq 
where $\sigma_{\rm T} \approx 6.65 \times 10^{-25}$ cm$^2$ is the Thomson cross section.

\begin{figure} 
\includegraphics[width = 85mm]{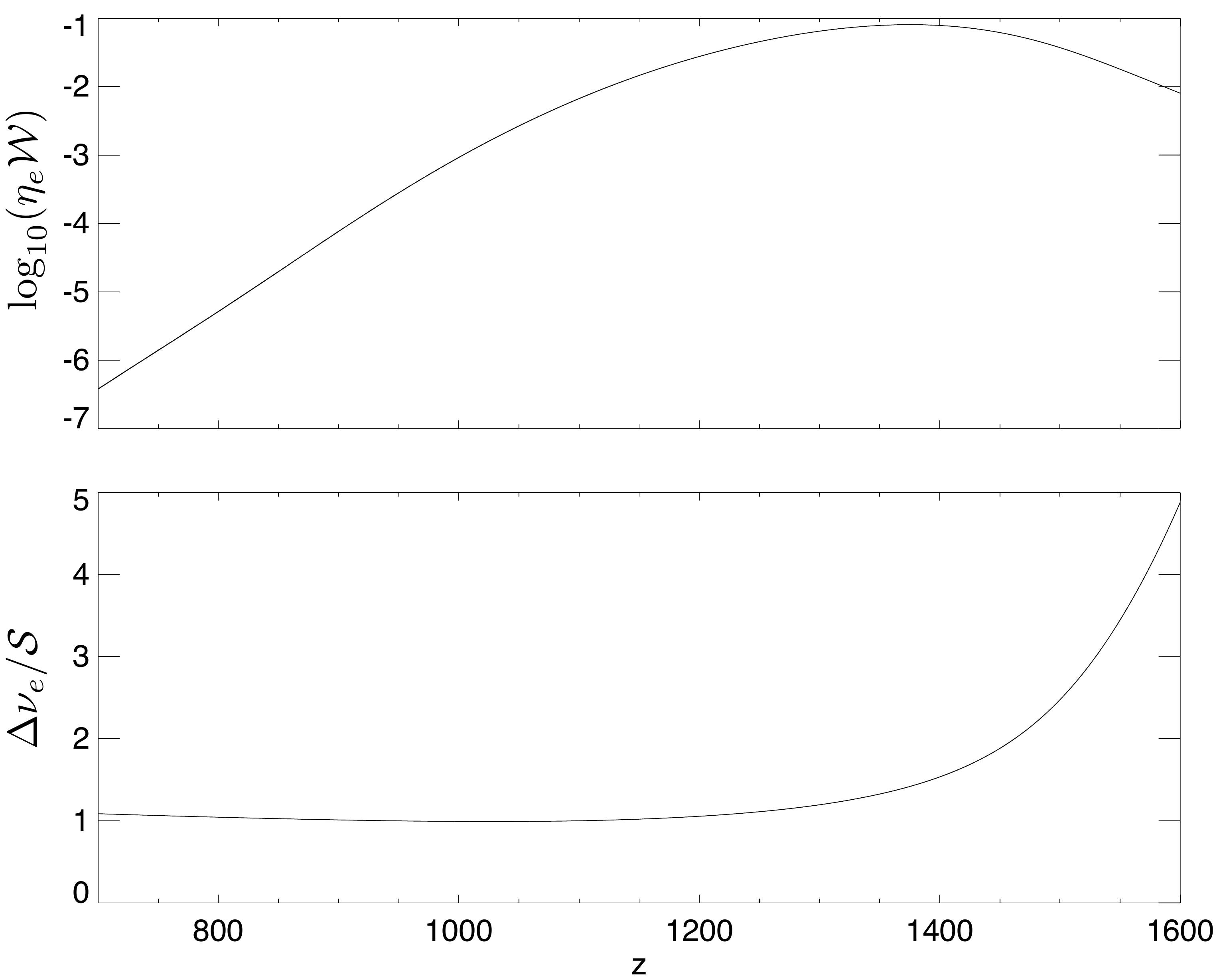}
\caption{\emph{Top panel}: characteristic number of electron scatterings within the region where the Ly$\alpha$ line is optically thick for true absorption. \emph{Bottom panel}: ratio of the characteristic frequency shift during a Thomson scattering event to the characteristic width over which the Ly$\alpha$ line is smoothed out by frequent resonant scatterings.}
\label{fig:thomson widths}
\end{figure}

Thomson scatterings can affect the recombination history if they take place within the characteristic width $\mathcal{W}$ over which the Lyman-$\alpha$ line is optically thick for true absorption. We show in Fig.~\ref{fig:thomson widths} the mean number of Thomson scatterings within a detuning $\mathcal{W}$ of line center, $\eta_e \mathcal{W}$. We see that it peaks at $\sim 0.08$ for $z \sim 1375$, and remains above 0.001 for $z \gtrsim 1000$, which suggests that Thomson scattering is potentially important at the subpercent level and should be carefully accounted for.

The rate of change of the number of photons per unit frequency per hydrogen atom due to Thomson scattering, neglecting stimulated scatterings, is:
\beq
\dot{\N}_{\nu}\big{|}_{\rm T} = N_{\rm H} x_{\rm e} \sigma_{\rm T} c \left[ - \N_{\nu} + \int \N_{\nu'} R_{\rm T}(\nu' \rightarrow \nu) \rmd \nu' \right],
\label{eq:rad_trans_T}
\eeq
where $R_{\rm T}(\nu' \rightarrow \nu)$ is the electron scattering kernel. 

If the radiation field is smooth on the scale of a characteristic frequency shift during a scattering $\Delta \nu_{\rme} \equiv \nu_{\Ly \alpha} \sqrt{2 k T_{\rm m}/m_e c^2}$, then the integral operator for electron scattering can be approximated by a Fokker-Planck operator, accounting for diffusion and drift in frequency space with rates \cite{Sazonov_Sunyaev}:
\barr
\frac{\rmd \langle \Delta \nu^2 \rangle}{\rmd t} &=& N_{\rm H} x_{\rm e} \sigma_{\rm T} c ~ \nu^2 \frac{2 k T_{\rm m}}{m_{\rme}c^2} \label{eq:diffusion}\\
\frac{\rmd \langle \Delta \nu \rangle}{\rmd t} &=& N_{\rm H} x_{\rm e} \sigma_{\rm T} c ~ \nu \frac{4 k T_{\rm m} - h \nu}{m_{\rme}c^2}.\label{eq:drift}
\earr
The corresponding Fokker-Planck equation is known as the Kompaneets equation:
\barr
\dot{\N_{\nu}}\big{|}_{\rm T}^{\rm FP} &=& N_{\rm H} x_{\rm e} \sigma_{\rm T} c \frac{k T_{\rm m}}{m_{\rme} c^2}\nonumber\\
&\times& \frac{\partial}{\partial \nu}\Bigg{\{}\nu^4 \left[ \frac{\partial}{\partial \nu}\left(\frac{\N_{\nu}}{\nu^2}\right) + \frac{h}{k T_{\rm m}} \frac{\N_{\nu}}{\nu^2}\right]\Bigg{\}}\label{eq:Kompaneets}
\earr
This is the approximation that was made in Ref.~\cite{CS09c}. However, due to the small mass of the electron, the characteristic frequency shift during an electron scattering event $\Delta \nu_{\rme }$ can be larger than the characteristic width over which the radiation field changes in the vicinity of the line. This characteristic width is of order $\mathcal{S}$ (which is $\gtrsim \mathcal{W}$ at all times, see Fig.~\ref{fig:WS.Ly.alpha}), set by frequency diffusion due to resonant scattering near line center (see end of Section \ref{section:rad transfer}). We see from Fig.~\ref{fig:thomson widths} that $\Delta \nu_e \geq \mathcal{S}$ at all times, and therefore electron scattering cannot be considered as a diffusive process and the Kompaneets equation is not valid in this context. 

We have implemented the correct integral scattering kernel given by Eq.~(\ref{eq:rad_trans_T}) in the Lyman-$\alpha$ transfer code developed in Ref.~\cite{Hirata_Forbes}. Accurate approximate expressions of the electron scattering kernel $R_{\rm T}(\nu' \rightarrow \nu)$ are given in Ref.~\cite{Sazonov_Sunyaev}. For the purpose of our calculation, we only need the kernel calculated in the non-relativistic limit adequate here, for a dipolar angular distribution \cite{Hummer_Mihalas67}. We set 
\barr
&&R_{\rm T}(\nu' \rightarrow \nu) + R_{\rm T}(\nu \rightarrow \nu') = \frac{2}{\Delta\nu_e} \mathcal{R}\left(\frac{\nu - \nu'}{\Delta \nu_e}\right),
\earr
where the dimensionless kernel $\mathcal{R}$ is given by\footnote{There is a typo in Ref.~\cite{Hummer_Mihalas67}: erf should be erfc.} \cite{Hummer_Mihalas67, Sazonov_Sunyaev}:
\barr
\mathcal{R}(\beta) &=& \frac1{10 \sqrt \pi} \left[11 + 4 \beta^2 + \frac12 \beta^4\right]\exp\left(-\frac{\beta^2}4\right)\nonumber\\
 &-& \frac14 \left[3 +  \beta^2 + \frac1{10}\beta^4\right] |\beta| ~ \textrm{erfc}\left(\frac{|\beta|}2\right), \label{eq:Thomson R unitless}
\earr
We moreover require that detailed balance is satisfied, i.e. that the Planck spectrum $\N_{\nu} \propto \nu^2 \rme^{- h\nu/(k T_{\rm m})}$ (in the limit $h \nu \gg k T_{\rm m}$ valid here) is preserved by imposing:
\beq
\frac{R_{\rm T}(\nu \rightarrow \nu')}{R_{\rm T}(\nu' \rightarrow \nu)} = \frac{\nu'^2}{\nu^2}\exp\left[\frac{h (\nu - \nu')}{k T_{\rm m}}\right].
\eeq
When evolving the number of photons per H nucleus per frequency bin in the $i^{\rm th}$ frequency bin, $N_i = \N_{\nu_i}\nu_i \Delta \ln \nu$, the radiative transfer code uses a backward Euler method which requires inverting the matrix equation:
\beq
M_{ij}(t + \Delta t) N_j(t + \Delta t) = N_i(t). 
\eeq
The matrix to be inverted, $\mathbf{M}$, is tridiagonal in the case where only absorption, emission, and resonant scattering (described by a Fokker-Planck operator) are present. Thomson scattering breaks this tridiagonality, which renders the system prohibitively time-consuming to invert ($\mathbf{M}$ is a 801$\times$801 matrix in our lowest resolution run). However, we can use the fact that Thomson scattering is only a perturbation to the radiative transfer equation. Therefore, $\mathbf{M}= \mathbf{M}_0 + \delta \mathbf{M}$, where $\mathbf{M}_0$ is an easily invertible tridiagonal matrix, and the perturbation $\delta \mathbf{M}$ due to Thomson scattering is such that its eigenvalues are always small compared to those of $\mathbf{M}_0$. We can therefore invert the perturbed matrix using the expansion:
\barr
\left(\mathbf{M}_0 + \delta \mathbf{M}\right)^{-1} &=& \mathbf{M}_0^{-1} - \mathbf{M}_0^{-1}(\delta \mathbf{M})\mathbf{M}_0^{-1} \nonumber\\
&+& \mathbf{M}_0^{-1}(\delta \mathbf{M})\mathbf{M}_0^{-1}(\delta \mathbf{M})\mathbf{M}_0^{-1}- ...
\earr 
We find that the second order of the expansion is usually sufficient, with a maximum change of the net decay rate in the line of $1.5 \times 10^{-5}$ between the first and second order.

\begin{figure} 
\includegraphics[width = 85mm]{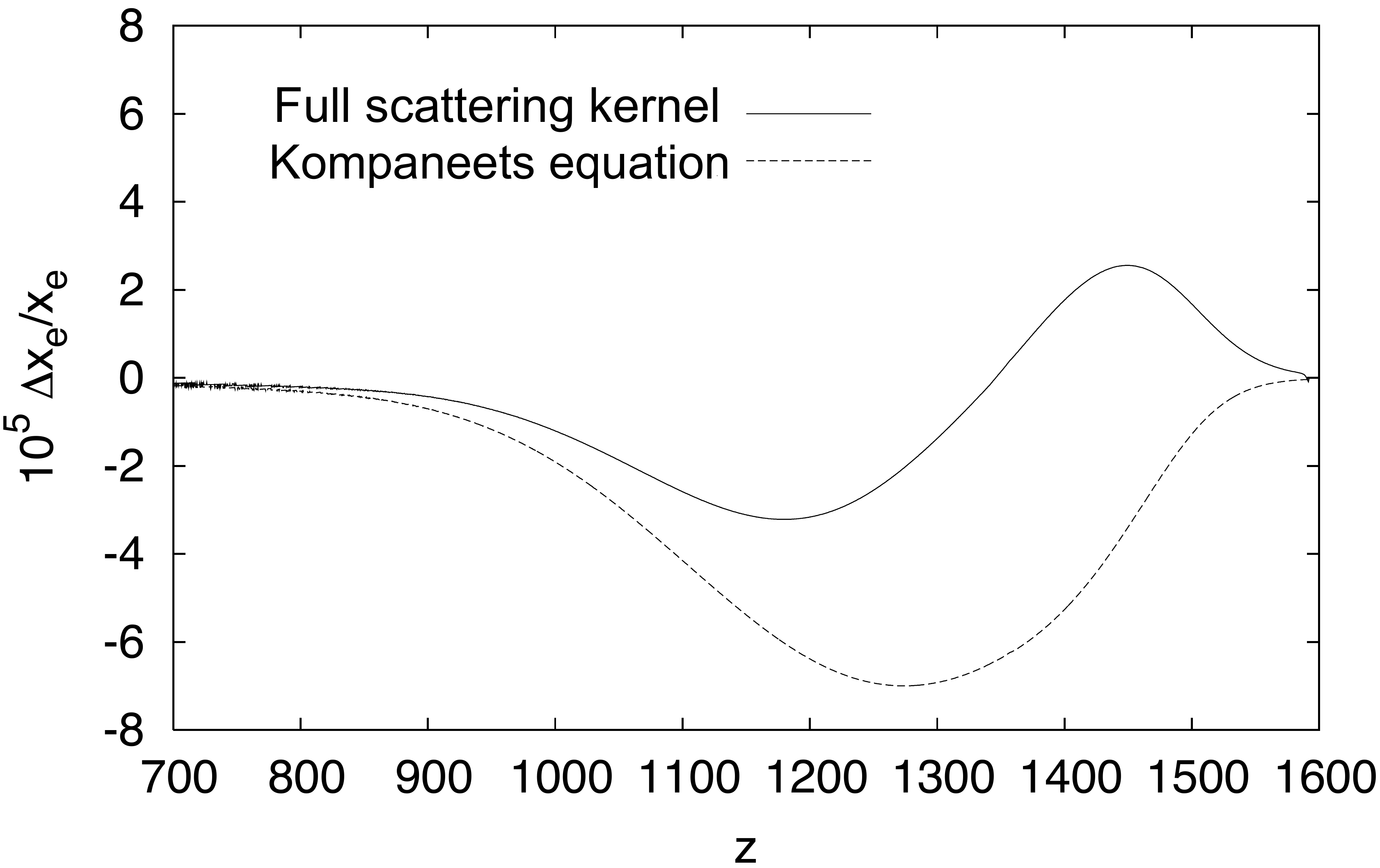}
\caption{Changes to the recombination history due to Thomson scattering.}
\label{fig:thomson Dxe}
\end{figure}

We show the resulting changes in the free electron fraction in Fig.~\ref{fig:thomson Dxe}. We can see that at early times, $z \gtrsim 1350$, Thomson scattering \emph{delays} recombination. Indeed, the relatively large frequency changes during electron scatterings allow photons to be moved from the red side of the line to the blue side of the line, and vice versa. Because of the large jump in photon occupation number across the line, the net photon flux is from the red side to the blue side. As a consequence, some escaping photons are reinjected into the line, where they can be absorbed, which decreases the escape rate and delays recombination. At later times, this effect is not so important as the radiation profile becomes smoother ($\Delta \nu_e \sim \mathcal{S}$, see Fig.~\ref{fig:thomson widths}). The systematic frequency loss during scattering event due to electron recoil starts to dominate, and Thomson scattering helps photons escaping out of the line and speeds up recombination.

 For comparison, we have also implemented the Kompaneets equation (\ref{eq:Kompaneets}), in a similar fashion as resonant scattering (see Ref.~\cite{Hirata_Forbes} for details on the implementation). We can see that using the Kompaneets equation does not represent accurately the physics of Thomson scattering, as it cannot capture the large frequency shifts at early times. The error in the correction is of order the correction itself, and it has the wrong sign at early times. However, the basic conclusion reached in Ref.~\cite{CS09c} remains valid: Thomson scattering can indeed be safely ignored during cosmic hydrogen recombination, since it leads to corrections to the ionization fraction of at most $\Delta x_e/x_e \sim \pm 3 \times 10^{-5}$.

\subsection{Interaction with the Deuterium Ly$\alpha$ line} \label{section:deuterium}

\subsubsection{Motivations}

A second radiative transfer effect associated with the Lyman-$\alpha$ transition is the interaction of the hydrogen and deuterium lines.
Due to the slightly larger reduced mass of deuterium, the Lyman-$\alpha$ frequency in deuterium $\nu_{\rm D} \equiv \nu_{\Ly \alpha}(\rm D)$ is shifted
to a higher frequency than that of hydrogen, $\nu_{\rm H} \equiv \nu_{\Ly \alpha}(\rm H)$. The relative shift, to first order in $m_{\rme}/m_{\rm p}$,
and with $m_{\rm D^+} \approx 2 m_{\rm p}$, is:
\beq
\frac{\nu_{\rm D} - \nu_{\rm H}}{\nu_{\rm H}} \approx \frac{m_{\rme}}{2 m_{\rm p}} \approx 2.7 \times 10^{-4}.
\eeq
This separation is $\sim 10$ times the Doppler width of the hydrogen line so the D Ly-$\alpha$ line center lies in the blue damping wing of the H
Ly-$\alpha$ line. Despite the tiny fractional abundance of primordial deuterium $x_{\rm D} = 2.87^{+0.22}_{-0.21} \times 10^{-5}$ \cite{Deuterium},
the D Ly-$\alpha$ line is still optically thick during cosmological hydrogen recombination, $\tau_{\textrm{D,Ly}\alpha} \approx x_{\rm D}
\tau_{\rm{H,Ly} \alpha}\sim 10^2-10^4$. This has motivated the authors of Ref.~\cite{Kholupenko_Deuterium} to consider the possible screening of radiation
incoming into the H Ly$\alpha$ line by the optically thick, bluer D Ly$\alpha$ line.  However, due to the optically thick nature of the H Ly$\alpha$
damping wings, the deuterium problem is more complicated than a simple Ly$\alpha$(D)$\rightarrow$Ly$\alpha$(H) feedback prescription, as pointed out in Ref.~\cite{Kholupenko_Deuterium}.
In this section, we set up the problem of the interaction of the H and D Ly$\alpha$ lines, and show that there is no significant effect on the
recombination history.

\subsubsection{Spectral distortions caused by deuterium}

Our first step is to understand the physical mechanism of deuterium recombination.  The rates of D and H recombination are tied together via the
charge-exchange reaction
\beq
\textrm{D}^+ + \textrm{H}(1s) \leftrightarrow \textrm{D}(1s) + \textrm{H}^+,
\label{eq:D-eq}
\eeq
which has a forward rate coefficient of order $\sim 10^{-9}\,$cm$^3\,$s$^{-1}$ \cite{Stancil_Deuterium}; at recombination-era densities of $\sim
500\,$cm$^{-3}$ this implies an equilibrium timescale of $\sim 2\times 10^6\,$s, i.e. six orders of magnitude shorter than the recombination timescale itself.  Thus to a very good approximation, the deuterium ionization fraction tracks that of hydrogen:
\beq
\frac{x_{\textrm{D}^+}}{x_{\textrm{D}(1s)}} \approx \frac{x_{\textrm{H}^+}}{x_{\textrm{H}(1s)}} \rme^{-\frac{\Delta E_{\rm I}}{k T_{\rm m}}} \approx
\frac{x_e}{1- x_e} \rme^{-\frac{\Delta E_{\rm I}}{k T_{\rm m}}},
\label{eq:D_abundance}
\eeq
where the difference in ionization energies of deuterium and hydrogen, $\Delta E_{\rm I}/k\approx 41$ K, is small compared to the matter
temperature during recombination.

Like hydrogen, deuterium may recombine radiatively and reach the ground state either via $2s\rightarrow 1s$ two-photon decay or Ly$\alpha$ escape; the
{\em net} rate of recombinations (H+D) is simply the sum of the two rates, and charge exchange (Eq.~\ref{eq:D-eq}) distributes the bound electrons
between H and D.  It is clear that the absolute rate of transitions to the ground state via the optically thin $2s\rightarrow 1s$ transition will be much smaller
for D than H because of its lower abundance; however for the Ly$\alpha$ channel in principle the rates could be comparable because the Ly$\alpha$(D)
optical depth is smaller than the Ly$\alpha$(H) optical depth by a factor of $\sim x_{\rm D}$, and hence the escape probability is enhanced by a
factor of $x_{\rm D}^{-1}$.  However, since Ly$\alpha$(D) is located blueward of Ly$\alpha$(H), all photons emitted in Ly$\alpha$(D)
will be re-absorbed in Ly$\alpha$(H); and all photons absorbed in the Ly$\alpha$(D) transition would have been absorbed anyway had the deuterium not been present.  Thus the {\em
net} (integrated over time) number of recombinations that proceed via Ly$\alpha$(D) escape is zero.  However, the re-absorption of the D photons is not instantaneous, so at
any given time the presence of deuterium causes an additional
distortion of the radiation field (on top of the distortion to the
blackbody spectrum due to H$(2p) \rightarrow \textrm H(1s)$ decays in the blue wing of H Ly$\alpha$). The number of additional distortion photons per hydrogen atom is
\beq
U = \int \frac{8\pi\nu^2}{c^3N_{\rm H}}[f_\nu^{(\rm H + D)}-
f_\nu^{(\rm H)}] ~\rmd\nu,
\label{eq:U-def}
\eeq
where $f_{\nu}^{(\rm H)}$ is the photon occupation number when only hydrogen is present, and $f_{\nu}^{(\rm H + D)}$ is its value when the presence of deuterium is accounted for.

Note that $U(t)\rightarrow 0$ at both early times (because the radiation field is thermal independently of the presence of deuterium) and late times (since after recombination is over, there are no more Ly$\alpha$ photons produced).

The correction to the rate of formation of ground-state atoms through the (H+D) $2p \rightarrow 1s$ channel is then equal to the rate of creation of distortion photons:
\beq
\Delta \dot x_{1s}|_{{\rm Ly}\alpha{\rm(D)}}  = - \Delta \dot x_e|_{{\rm Ly}\alpha{\rm(D)}} = \dot{U}.\label{eq:U.dotxe}
\eeq

\subsubsection{Analytic estimate for the number of spectral distortion photons}

We may now obtain an analytic estimate for $U(t)$. Following Hirata
\cite{Hirata_2photon} and Hirata \& Forbes \cite{Hirata_Forbes}, we make the change of variables
\beq
\nu \equiv \nu_{\rm H} + \frac{k T_{\rm r}}{h} y,
\eeq
and write 
\beq
f_{\nu} = \rme^{-\frac{h \nu}{k T_{\rm r}}} + \left(f_{\rm eq}^{\rm H}
  - \rme^{-\frac{h \nu_{\rm H}}{k T_{\rm r}}} \right) \Psi(y), \label{eq:fnu.change.var}
\eeq
where $f_{\rm eq}^{\rm H} \equiv x_{\textrm{H}(2p)} / \left(3 x_{\textrm{H}(1s)}\right)$.
We also define the dimensionless widths:
\barr
W &\equiv& \frac{h \mathcal{W}}{k T_{\rm r}}, \\
S &\equiv&  \left(\frac{h
\mathcal{S}}{k T_{\rm r}}\right)^3,\\
y_{\rm D} &\equiv& \frac{h(\nu_{\rm D} - \nu_{\rm H})}{k T_{\rm r}}.
\earr 
Taking $T_{\rm m} \approx T_{\rm r}$ and using the damping wing
approximation for the Voigt profile, the steady-state radiative
transfer equation in the blue wing of H Ly$\alpha$ takes the following
form in the absence of deuterium \cite{Hirata_Forbes}:
\barr
\frac{\rmd \Psi^{(\rm H)}}{\rmd y} &=& \frac{W}{y^2}\left(\rme^y
  \Psi^{(\rm H)} - 1\right) \nonumber \\
&-& \frac{\rmd}{\rmd
  y}\left[\frac{S}{y^2}\left(\frac{\rmd \Psi^{(\rm H)}}{\rmd y} +
    \Psi^{(\rm H)}\right)\right], \label{eq:Psi_H}
\earr
with boundary conditions $\Psi^{(\rm H)}(+\infty) = 0$ (neglecting Ly$\beta\rightarrow$ Ly$\alpha$ feedback in this problem for simplicity) and $\Psi^{(\rm
H)}(0) = 1$. The presence of deuterium modifies this equation by
adding a term due to absorption and emission by deuterium at $y = y_{\rm D}$:
\barr
\frac{\rmd \Psi^{(\rm H + D)}}{\rmd y} &=& \frac{W}{y^2}\left(\rme^y
  \Psi^{(\rm H+D)} - 1\right) \nonumber\\
&-& \frac{\rmd}{\rmd
  y}\left[\frac{S}{y^2}\left(\frac{\rmd \Psi^{(\rm H+D)}}{\rmd y} +
    \Psi^{(\rm H+D)}\right)\right] \nonumber\\
&+& \tau_{\rm D} \varphi_{\rm
  D}(y) \left(\rme^{y - y_{\rm D}}\Psi^{(\rm H + D)}(y) - \Psi_{\rm
    D}\right)  \label{eq:Psi_D}
\earr
where $\tau_{\rm D} \equiv \tau_{\rm D, \Ly \alpha}$, $\varphi_{\rm
  D}(y) \equiv h/k T_{\rm r}~\varphi_{\rm D}(\nu)$ is the
dimensionless emission profile in the deuterium
line, and we have used
\beq
\Psi_{\rm D} \equiv \frac{   f_{\rm eq}^{\rm D}
  - \rme^{-\frac{h \nu_{\rm D}}{k T_{\rm r}}}    }{  f_{\rm eq}^{\rm H}
  - \rme^{-\frac{h \nu_{\rm H}}{k T_{\rm r}}}  }, \label{eq:PsiD.def}
\eeq
with $f_{\rm eq}^{\rm D}\equiv x_{\textrm{D}(2p)} / \left(3
  x_{\textrm{D}(1s)}\right)$.

Several implicit assumptions went
into Eq.~(\ref{eq:Psi_D}).

First, we assumed complete redistribution
in the deuterium line (i.e. did not account for partial redistribution
during resonant scattering events). This approximation is well
justified since the deuterium line is dominated by its Doppler core: the damping wings are only marginally optically thick, and the
differential optical depth in the deuterium wings is always much
smaller than that of hydrogen anyway (see for example Fig.~7 of
Ref.~\cite{Kholupenko_Deuterium}). In this case complete redistribution is a good
approximation since both partial and complete redistribution have a similar characteristic frequency width, the Doppler width of the line.

In addition, we assumed that the Fokker-Planck
representation of the scattering operator (for scatterings by hydrogen atoms) remained valid in the
vicinity of the deuterium line. This can only be valid if the
radiation field remains smooth on the scale of a Doppler width, even
with the presence of the narrow optically thick deuterium line. This
is the case since $(\nu_{\rm D} - \nu_{\rm H})^3 \ll \mathcal{S}^3$ at
most times (see Fig.~\ref{fig:WS.Ly.alpha}, with $\nu_{\rm D} - \nu_{\rm H} \approx 10\times
\nu_{\rm H}\Delta_{\rm H}$). Therefore, according to the discussion in
Section~\ref{section:rad transfer}, diffusion occurs on a much shorter timescale than
redshifting, and photons emitted in deuterium Ly$\alpha$ decays at the
deuterium resonant frequency are rapidly redistributed on both sides
of the line by frequent scatterings by hydrogen atoms, smoothing out the radiation field
in  the vicinity of the D Ly$\alpha$ line.

Finally, one should in principle rescale the optical depth for hydrogen true absorption and scattering by a factor $(1 - x_{\rm D})$, in both the blue and red wings (in particular, change $W$ and $S$ to $(1 - x_{\rm D})W$ and $(1 - x_{\rm D}) S$ in Eq.~(\ref{eq:Psi_D})). However, Ly$\alpha$ decays in the damping wings only lead to a $\sim 1\%$ correction to the recombination history \cite{Hirata_2photon}. Therefore the correct rescaling would lead to a negligible $\mathcal{O}(10^{-7})$ correction, and we simply use $W,S$ in Eq.~(\ref{eq:Psi_D}) and neglect corrections to the decay rate in the red wing.
 
Our last approximation will consist in taking $\varphi_{\rm D}(y) =
\delta(y - y_{\rm D})$. Indeed, the radiation field will vary on a characteristic
scale $\mathcal{S}$, which is much larger than the
width of the Doppler-broadened deuterium line (see Fig.~\ref{fig:WS.Ly.alpha}; the Doppler width of the deuterium line is $\sqrt 2$ times smaller than that of the hydrogen line). The
presence of the optically thick deuterium line in the blue wing of H
Ly$\alpha$ therefore amounts to imposing an additional boundary
condition:
\beq
\Psi^{(\rm H + D)}(y_{\rm D}^-) = \Psi^{(\rm H + D)}(y_{\rm D}^+) =
\Psi_{\rm D}, 
\eeq
where the continuity of $\Psi^{(\rm H + D)}$ across the
deuterium line is insured by the very fast redistribution of emitted
photons in the frequency domain by the frequent hydrogen resonant
scatterings. Eq.~(\ref{eq:Psi_D}) is otherwise identical to
Eq.~(\ref{eq:Psi_H}) for $y > y_{\rm D}$ and $y < y_{\rm D}$. We define
\beq
\Delta \Psi_{\rm D} \equiv \Psi_{\rm D} - \Psi^{(\rm H)}(y_{\rm D}).\label{eq:DPsiD.def}
\eeq
By linearity, we have 
\beq
\Psi^{(\rm H + D)} - \Psi^{(\rm H)} = \Delta\Psi_{\rm D} \times \psi \label{eq:psi def}, 
\eeq
where the function $\psi(y)$ is the solution of the
second order linear homogeneous equation
\beq
\frac{\rmd \psi}{\rmd y} = \frac{W}{y^2}\rme^y
  \psi  - \frac{\rmd}{\rmd
  y}\left[\frac{S}{y^2}\left(\frac{\rmd \psi}{\rmd y} +
    \psi\right)\right] \label{eq:psi}
\eeq
with boundary conditions
\beq
\psi(+\infty) =  \psi(0) = 0 \ \ \ , \ \ \ \psi(y_{\rm D}) = 1.
\eeq
The number of additional distortion photons per hydrogen atom due the
presence of deuterium therefore becomes
\beq
U \approx \frac{8 \pi \nu_{\Ly \alpha}^2 k T_{\rm r}}{c^3 N_{\rm H} h}\left(f_{\rm eq}^{\rm H}
  - \rme^{-\frac{h \nu_{\rm H}}{k T_{\rm r}}} \right) \Delta \Psi_{\rm
D} \Delta \mathcal{I}, \label{eq:U.analytic}
\eeq
where
\beq
\Delta \mathcal{I}(W, S, y_{\rm D}) \equiv \int_0^{+\infty} \psi(y)
\rmd y.
\eeq
Our last step is to estimate $\Delta \Psi_{\rm D}$. The population of the $n=2$ shell of deuterium is
controlled by recombinations to the excited states and Lyman-$\alpha$
decays (as mentioned earlier, decays to the ground state through the
Lyman-$\alpha$ channel are dominant over two-photon decays from the
$2s$ state because of the relatively low optical depth of the D
Ly$\alpha$ line). We can obtain the contribution of the former with
a Peebles-like estimate \cite{Peebles}:
\barr
\dot{x}_{\textrm{D}(2p)}\big{|}_{\rm rec} &=& \frac34 \alpha_{\rm B}(T_{\rm m}) \Bigg{[}N_{\rm H} x_{\rme} x_{\rm D^+}\nonumber\\
&-& \frac13 \left(\frac{m_{\rme} k T_{\rm m}}{2 \pi \hbar^2} \right)^{3/2} \rme^{\frac{- h \nu_{\rm D}}{3 k T_{\rm m}}} x_{\textrm{D}(2p)}\Bigg{]},
\label{eq:D_rec}
\earr
where $\alpha_{\rm B}(T_{\rm m})$ is the case-B recombination
coefficient, for which we use the fitting function given in
Ref.~\cite{alphaB}. The contribution from the Ly$\alpha$ decays,
Eq.~(\ref{eq:net decay simple}), can be rewritten after some manipulations as:
\barr
&&\dot{x}_{\textrm{D}(2p)}\big{|}_{\Ly \alpha} = \frac{8 \pi H \nu_{\rm
    D}^3}{c^3 N_{\rm H}} \left(f_{\rm eq}^{\rm H}
  - \rme^{-\frac{h \nu_{\rm H}}{k T_{\rm r}}} \right) \nonumber \\
&&\times \int \tau_{\rm D} \varphi_{\rm
  D}(y) \left(\rme^{y - y_{\rm D}}\Psi^{(\rm H + D)}(y) - \Psi_{\rm
    D}\right) \rmd y. 
\earr
Using Eq.~(\ref{eq:Psi_D}), and taking the limit $\varphi_{\rm D}(y) \rightarrow
\delta(y - y_{\rm D})$ (i.e. assuming $\varphi_{\rm D}(y)$ has support only in $y_{\rm D} \pm \epsilon$, and taking the limit $\epsilon \rightarrow 0$) we obtain:
\barr
&&\int \tau_{\rm D} \varphi_{\rm
  D}(y) \left(\rme^{y - y_{\rm D}}\Psi^{(\rm H + D)}(y) - \Psi_{\rm
    D}\right) = \nonumber\\
&&\frac{S}{y_{\rm D}^2} \left[\frac{\rmd \Psi^{(\rm
      H+D)}}{\rmd y}\Big{|}_{y_{\rm D}^+} - \frac{\rmd \Psi^{(\rm
      H+D)}}{\rmd y}\Big{|}_{y_{\rm D}^-} \right], \label{eq:dot.xD2p}
\earr
where we used the continuity of $\Psi^{(\rm H + D)}$ across the
line (so the first derivative integrates to $\Psi^{(\rm H + D)}(y_{\rm D}+\epsilon) - \Psi^{(\rm H + D)}(y_{\rm D}-\epsilon) \rightarrow 0$), and its finiteness (so the integral of the term proportional to $W$ vanishes as $\epsilon \rightarrow 0$). The \emph{derivative} of $\Psi^{(\rm H + D)}$ may however be
discontinuous across the deuterium line because of the presence of
the delta function. Eq.~(\ref{eq:dot.xD2p}) simply states that the net rate of decays in the deuterium line is balanced, in steady-state, by the flux of photons due to
frequency diffusion across $\nu_{\rm D}$. Since there is no discontinuity of the derivative of $\Psi^{(\rm H)}$ at $y_{\rm D}$, we can replace $\Psi^{(\rm H +
  D)}$ by $\Psi^{(\rm H + D)}- \Psi^{(\rm H)} = \Delta \Psi_{\rm D} \times
\psi$ in Eq.~(\ref{eq:dot.xD2p}). Our final expression for the net
rate of Ly$\alpha$ absorptions is therefore
\beq
\dot{x}_{\textrm{D}(2p)}\big{|}_{\Ly \alpha} = -\frac{8 \pi H \nu_{\rm
    D}^3}{c^3 N_{\rm H}} \left(f_{\rm eq}^{\rm H}
  - \rme^{-\frac{h \nu_{\rm H}}{k T_{\rm r}}} \right) \Delta \Psi_{\rm
  D} \Delta \Xi, \label{eq:dot.xD2p.Lya}
\eeq
where
\beq
\Delta \Xi(W, S, y_{\rm D}) \equiv \frac{S}{y_{\rm D}^2}\left[\frac{\rmd \psi}{\rmd y}\Big{|}_{y_{\rm D}^-} - \frac{\rmd \psi}{\rmd y}\Big{|}_{y_{\rm D}^+} \right] > 0.
\eeq
Transition rates into and out of the $2p$ state of deuterium are many
orders of magnitude larger than the overall
recombination rate which is of order of the Hubble rate. The
population of the $2p$ state, and therefore $\Delta \Psi_{\rm D}$, can
be solved for in the steady-state approximation, setting
\beq
0 \approx \dot{x}_{\textrm{D}(2p)} = \dot{x}_{\textrm{D}(2p)}\big{|}_{\rm rec} +
\dot{x}_{\textrm{D}(2p)}\big{|}_{\Ly \alpha}. \label{eq:ss.deuterium}
\eeq
We first express $x_{\textrm{D}(2p)}$ in terms of $\Delta \Psi_{\rm D}$, using Eqs.~(\ref{eq:fnu.change.var}), (\ref{eq:PsiD.def}) and (\ref{eq:DPsiD.def}):
\barr
&&x_{\textrm{D}(2p)} = 3 x_{\textrm{D}(1s)} \times  \nonumber \\
&& \left[ \rme^{- \frac{h
      \nu_{\rm D}}{k T_{\rm r}}} + \left(f_{\rm eq}^{\rm H}
  - \rme^{-\frac{h \nu_{\rm H}}{k T_{\rm r}}} \right) \left(\Psi^{(\rm
  H)}(y_{\rm D}) + \Delta \Psi_{\rm D}\right) \right].
\earr
We also need to compute $\Psi^{(\rm H)}(y_{\rm D}), \Delta \Xi$ and $\Delta \mathcal{I}$ numerically as a function of $W, S, y_{\rm D}$. We do so by solving Eqs.~(\ref{eq:Psi_H}) and (\ref{eq:psi}) with a shooting method (see Appendix A of Ref.~\cite{Hirata_Forbes}). Using Eq.~(\ref{eq:D_abundance}) for $x_{\rm D^+}$ and
replacing $x_{\textrm{D}(2p)}$ by the above expression in Eq.~(\ref{eq:D_rec}), we can then solve for $\Delta \Psi_{\rm D}$ for a given ``order zero''
recombination history (i.e. computed without the deuterium
correction) and the corresponding parameters $W(z), S(z), f_{\rm eq}^{\rm H}(z)$. We use the output of our EMLA code \cite{EMLA} to generate such a history. 

\subsubsection{Results and discussion}

We obtain a very small relative distorsion $\Delta \Psi_{\rm D}$, with a maximum value of $\sim 10^{-4}$ at $z \sim 1600$, falling below $10^{-6}$ for $z \lesssim 1400$. As a result, the number of distortion photons per hydrogen atom, $U$, given by Eq.~(\ref{eq:U.analytic}) is also extremely small, peaking at $\sim 3 \times 10^{-9}$ at $ z \sim 1500$, and D Ly$\alpha$ decays have virtually no effect on the recombination dynamics.

To better understand the magnitude of the effect, it is more enlightening at this point to write some approximate expressions in the case relevant here.

We start by rewriting the net decay rate in the deuterium Ly$\alpha$ line, Eq.~(\ref{eq:dot.xD2p.Lya}), in a more familiar form, using the definitions for $\Delta \Psi_{\rm D}$, $\Psi_D$ and $\Psi^{\rm H}(y)$ Eqs.~(\ref{eq:DPsiD.def}), (\ref{eq:PsiD.def}) and (\ref{eq:fnu.change.var}):
\barr
\dot{x}_{\textrm{D}(2p)}\big{|}_{\Ly \alpha} &=& -\Delta \Xi \frac{8 \pi H \nu_{\rm
    D}^3}{3 c^3 N_{\rm H} x_{\textrm{D}(1s)}}\nonumber\\
&\times& \left(x_{\textrm{D}(2p)} - 3 x_{\textrm{D}(1s)}f_{\nu_{\rm D}}^+\right), \label{eq:dotxD_new}
\earr
where $f_{\nu_{\rm D}}^+ \equiv f^{\rm H}_{\nu_{\rm D}}$. Without the factor $\Delta \Xi$, this is just the usual Sobolev decay rate, Eq.~(\ref{eq:Sobolev rate bis}) which is the rate at which photons emitted in the line redshift across the line.

Now, as can be seen from Fig.~\ref{fig:WS.Ly.alpha}, at most relevant times, $W^3 \ll S$. Moreover, as the separation of the deuterium and hydrogen lines is approximately 10 Doppler widths, we also have $y_{\rm D}^3 \ll S$. In these limits, one can show that $\Delta \Xi$ takes on the approximate form:
\beq
\Delta \Xi \approx 3 \frac S{y_{\rm D}^3} = 3 \left(\frac{\mathcal{S}}{\nu_{\rm D} - \nu_{\rm H}}\right)^3 \gg 1.
\eeq 
Using Eq.~({\ref{eq:S.meaning}}), we see that $\Delta \Xi \sim \Delta t_{\rm redshift}/\Delta t_{\rm diff}$. The meaning of Eq.~(\ref{eq:dotxD_new}) is now clearer: the net D $2p\rightarrow 1s$ decay rate is given by the rate at which photons \emph{diffuse} out of the deuterium line due to scattering by hydrogen atoms, rather than the rate at which they redshift out of the line as would be prescribed by the Sobolev approximation. The fast frequency diffusion rate near $\nu_{\rm D}$ therefore enhances the decay rate in the line. Note that the decay rate \emph{per deuterium atom} is already enhanced by a factor of $x_{\rm D}^{-1}$ relative to the equivalent rate in the hydrogen line, per hydrogen atom. Therefore the very fast $2p \rightarrow 1s$ decays bring the deuterium $2p$ to $1s$ ratio towards near equilibrium with the local radiation field, $f_{\nu_{\rm D}}^{\rm H}$ (if it were not for the distortions due to the optical thickness of the blue wing of hydrogen, the $2p$ state would therefore rather be near Boltzmann equilibrium with the ground state than near Saha equilibrium as it is the case for hydrogen). 

In the net recombination rate to the D($2p$) state, Eq.~(\ref{eq:D_rec}), we can therefore approximate $x_{\textrm{D}(2p)} \approx 3 x_{\textrm{D}(1s)} f_{\nu_{\rm D}}^{\rm H} \approx \rme^{- h(\nu_{\rm D} - \nu_{\rm H})/kT_{\rm r}}x_{\rm D} x_{\textrm{H}(2p)}$. The second approximation stems from the fact that $(\nu_{\rm D} - \nu_{\rm H})^3 \ll \mathcal{S}^3$. With this approximation, and using Eq.~(\ref{eq:D_abundance}), we obtain:
\beq
\dot{x}_{\textrm{D}(2p)}\big{|}_{\rm rec} \approx x_{\rm D} \dot{x}_{\textrm{H}(2p)}\big{|}_{\rm rec} 
\eeq
We also rewrite the net decay rate in the D Ly$\alpha$ line, Eq.~(\ref{eq:dot.xD2p.Lya}), as follows:
\barr
\dot{x}_{\textrm{D}(2p)}\big{|}_{\textrm{Ly}\alpha} &=& \dot{x}_{\textrm{H}(2p)}\big{|}_{\textrm{Ly}\alpha} \times \Delta \Psi_{\rm D} \Delta \Xi\nonumber\\
&\approx& - \dot{x}_{\textrm{H}(2p)}\big{|}_{\rm rec} \times \Delta \Psi_{\rm D} \Delta \Xi,
\earr
where in the second line we used the steady-state approximation for the population of the H$(2p)$ state. Setting $\dot{x}_{\textrm{D}(2p)} \approx 0$, we therefore obtain:
\beq
\Delta \Psi_{\rm D} \approx \frac{x_{\rm D}}{\Delta \Xi}.
\eeq
Now, roughly speaking, for $y_{\rm D} \ll W, S$, the integral $\Delta \mathcal{I} = \int \psi(y) \rmd y$ is of the same order of magnitude as the integral $\int \Psi^{\rm H}(y) \rmd y$ (this can be checked numerically). This is only very approximate but allows us to rewrite
\beq
\frac{8 \pi \nu_{\Ly \alpha}^2 k T_{\rm r}}{c^3 N_{\rm H} h}\left(f_{\rm eq}^{\rm H}
  - \rme^{-\frac{h \nu_{\rm H}}{k T_{\rm r}}} \right) \Delta \mathcal{I} \sim x_+,
\eeq
where $x_+$ is the number of distortion photons per hydrogen atoms in the blue wing of Ly$\alpha$, due to two-photon transitions and diffusion in hydrogen (see Eq.~(69) of Ref.~\cite{Hirata_Forbes}). 

We can now finally obtain an approximate expression for the number of distortion photons due to the presence of deuterium:
\beq
U \sim \frac{x_{\rm D}}{\Delta \Xi} x_+. \label{eq:U-approx}
\eeq
The distortion in the blue wing of Lyman-$\alpha$ due to hydrogen two-photon decays and diffusion itself leads to $\mathcal{O}(1\%)$ corrections to the recombination history \cite{Hirata_2photon, Hirata_Forbes}. The distortion due to the presence of deuterium is much smaller due to the very fast D Ly$\alpha$ decays, which bring the D($2p$)/D($1s$) ratio very close to equilibrium with the radiation field at the frequency $\nu_{\rm D}$. As explained above, this fast decay rate is due to a relatively small optical depth in the deuterium Lyman line (smaller by a factor $x_{\rm D}$ than that of hydrogen), and is further enhanced by the large diffusion rate near $\nu_{\rm D}$, translating to the large $\Delta \Xi$ in the denominator of Eq.~(\ref{eq:U-approx}).

We therefore conclude that the effect of deuterium on the recombination history is at most $\mathcal{O}(10^{-7})$ (due to the rescaling of the optical depth of hydrogen absorption and scattering in the damping wings, changing $x_{+}$ by factors of order $x_{\rm D}$). Note that accounting for the presence of deuterium also induces changes to the background expansion of the order of $x_{\rm D} \sim 10^{-5}$. Moreover, depending on how the ionization fraction is normalized (with respect to total H+D of just H), there can be ambiguities of $\sim 10^{-5}$ in $x_e$. These modifications to the recombination histories can however be safely ignored given the level of accuracy required by \emph{Planck}.

\section{Higher order, non-overlapping Lyman lines ($2 \le n \lesssim 23$)}\label{section:resonant scat}

\subsection{List of efficient Lyman transitions}

One of the strengths of the EMLA formulation \cite{EMLA} is that it only requires the computations of effective transition rates for a small set of ``interface'' states. These ``interface'' states should in principle be $2s, 2p$, and all the higher lying $p$ states. However, as exposed in Ref.~\cite{EMLA}, only the lowest few Lyman transitions significantly contribute to the overall recombination rate, and one can neglect higher order Lyman transitions without loss of accuracy. The aim of this section is to verify this statement quantitatively.

For that purpose, we computed a series of recombination histories with \textsc{RecSparse} \cite{Grin_Hirata}, with $n_{\max} = 30$, artificially setting the Lyman transition rates to zero above a fiducial transition, and compared them to the exact calculation where all Lyman transitions are included up to $n_{\max}$. Both calculations included feedback between neighboring Lyman lines with non-vanishing transition rates, and in both cases the photon occupation number incoming on the highest line considered was taken to be a blackbody. We find that neglecting Ly$\gamma$ and above leads to changes to the recombination history $|\Delta x_{e}/x_e|$ of at most $8 \times 10^{-5}$, neglecting Ly$\delta$ and above leads to changes of at most $10^{-5}$, and neglecting Ly$\epsilon$ and above leads to changes of at most $3 \times 10^{-6}$. We therefore conclude that it is sufficient to only include the Ly$\alpha$, $\beta$, and $\gamma$ transitions when computing recombination histories accurate to the level of 0.01\%.

\subsection{Resonant scattering in the low lying Lyman lines}
The Lyman-$\alpha$ line has a particularly high resonant scattering probability, as atoms in the $2p$ state can leave the state only through spontaneous Ly-$\alpha$ decay, or through absorption of a CMB photon. There is no lower energy state to spontaneously decay to other than the ground state. For higher order Lyman transitions, though, there are multiple allowed channels out of the $np$ state, $np\rightarrow n'l$, with $1<n'<n$ and $l = 0,2$. Therefore the scattering probability defined in Eq.~(\ref{eq:pscat}) becomes comparable to the absorption probability, $\pscn \sim \pabn$ (in vacuum, $p_{\rm ab}^{\rm Ly \beta} \approx 0.12$ and $p_{\rm ab}^{\rm Ly \gamma} \approx 0.16$). From Eqs.~(\ref{eq:Wn}), (\ref{eq:Sn}), we can see that in that case 
\beq
\left(\frac{\mathcal{S}_n}{\nu_n \Delta_{\textrm{H}}}\right)^3 \sim \frac{\mathcal{W}_n}{\nu_n \Delta_{\textrm{H}}}. \label{eq:WsimS}
\eeq
The wings of the Ly$\beta$ and Ly$\gamma$ lines are optically thick for true absorption (this is the case for all Ly-$n$ lines with $n \lesssim 13$). This implies (see Eq.~(\ref{eq:Wn}) and corresponding discussion):
\beq
1 \ll \frac{\mathcal{W}_n}{\nu_n \Delta_{\textrm{H}}} \ll \left(\frac{\mathcal{W}_n}{\nu_n \Delta_{\textrm{H}}} \right)^3,
\eeq
where the the second inequality is a consequence of the first one. From Eq.~(\ref{eq:WsimS}), we therefore obtain that for the low-lying Lyman lines above Ly$\alpha$, $\left(\mathcal{S}_n/\mathcal{W}_n\right)^3 \ll 1$. This means that frequency diffusion is efficient only on a small fraction of the width over which the line is optically thick to true absorption.

There is not a simple relationship between the ratio $\left(\mathcal{S}_n/\mathcal{W}_n\right)^3$ and the impact of frequency diffusion on the Ly-$n$ decay rate. However, the effect of frequency diffusion clearly increases with this ratio. Radiative transfer computations including frequency diffusion were carried out for the Lyman-$\alpha$ line, which showed that frequency diffusion leads to corrections of a few percents to the net decay rate in Ly$\alpha$ \cite{Hirata_Forbes, CS09c}. As shown in Fig.~\ref{fig:WS.Ly.alpha}, $\left(\mathcal{S}_2 /\mathcal{W}_2\right)^3 \gg 1$ at most times for the Ly$\alpha$ line, in contrast with what we just showed for the higher-order lines. Therefore, we can expect that frequency diffusion would lead to corrections of much less than a percent to the net decay rate in the Ly$\beta$ and Ly$\gamma$ lines. Since Ly$\beta$ decays themselves contribute of the order of a percent only to the overall recombination rate, sub-percent corrections to their rate can therefore be safely neglected at the level of accuracy required.

As a conclusion, resonant scattering in Ly$\beta$ and higher order Lyman lines does not affect the recombination dynamics to any significant level.

\section{Overlap of the high-lying Lyman lines ($n \gtrsim 24$)}\label{section:overlap}
\subsection{Motivations}

In the previous section we have shown that Lyman transitions above Ly$\gamma$ do not affect the recombination history to a significant level and can be ignored. However, this relied on using the Sobolev approximation for the net decay rate out of the optically thick Lyman lines. As we show below, high-lying Lyman lines overlap with each other, or even with the continuum, which breaks assumption (i) of the Sobolev approximation that each line is isolated. In this section we address the consequences of this feature. 

For $n \gtrsim 13$, the Lyman lines are dominated by their Doppler core (the damping wings become optically thin). The condition for the two neighboring Ly-$(n+1)$ and Ly-$n$ lines to overlap is therefore that their separation $\nu_{n+1} - \nu_n \sim 2 \nu_c / n^3$ becomes of order the Doppler width of an individual line, $\nu_n \Delta_{\rm H} \sim \nu_c \Delta_{\rm H}$. Overlap therefore occurs for
\beq
n \gtrsim n_{\rm ov} \equiv \left(\frac{2} {\Delta_{\rm H}}\right)^{1/3} \approx 44 \left(\frac{T_{\rm m}}{3000 ~\rm K}\right)^{-1/6}.\label{eq:n2}
\eeq
For even higher order lines, the separation with the continuum $\nu_c - \nu_n = \nuc / n^2$ becomes of the order of a Doppler width. This occurs for 
\beq
n \gtrsim n_{\rm ov,c} \equiv \Delta_{\rm H}^{-1/2} \approx 206 \left(\frac{T_{\rm m}}{3000 K}\right)^{-1/4}.
\eeq

Intuitively, one can expect that line overlap amounts to adding new transitions $R_{np \rightarrow n'p}^{(\rm ov)}$ between high-lying $p$ states, as photons emitted in the $np\rightarrow 1s$ transition can be re-absorbed immediately in a neighboring $1s \rightarrow n'p$ transition. More importantly, overlap with the continuum provides an additional recombination pathway. Direct recombinations to the ground state are usually considered as highly inefficient as the resulting emitted photons can immediately ionize neutral hydrogen atoms in their ground state. Chluba \& Sunyaev \cite{CS07} have considered the possibility of continuum escape (similar to Lyman-$\alpha$ escape), and have shown that it leads to negligible corrections to the ionization history $\Delta x_e/x_e \sim 10^{-6}$. If overlap of the highest Lyman lines with the continuum is accounted for, it becomes possible for free electrons and protons to successfully recombine to the ground state of hydrogen, if the emitted photon subsequently excites another atom to a high-lying $p$ state rather that ionizing it. The recombination event $e^- + p \rightarrow 1s + \gamma$ immediately followed by the absorption event $1s + \gamma \rightarrow np$ thus corresponds to an additional, indirect recombination event to the $np$ state, to which a coefficient $\alpha^{(\rm ov)}_{np}$ can be associated. The reverse process corresponds to an additional photoionization rate, $\beta_{np}^{(\rm ov)}$.

Modern recombination codes account for the excited states of hydrogen up to extremely high principal quantum number $n_{\max} \gtrsim 200$ \cite{Grin_Hirata, Chluba_Vasil}, and recently up to $n_{\max} = 500$ \cite{EMLA}. It is therefore important to quantify the impact of line overlap on the recombination history. In what follows we develop a formalism that generalizes the Sobolev escape probability method, and accounts for the overlap of the high lying Lyman lines. 

We start by evaluating the effect of photoionization and recombinations from and to the ground state on the radiation field and providing the relevant equations.

\subsection{Photoionization and recombination from and to the ground state}\label{section:photoionization}

The frequency-dependent photoionization cross-section from the ground state is, in the atom's rest frame:
\beq
\sigma(\nu) = \sigma_0 ~g\left(\frac{\nu}{\nuc}\right),
\eeq
where 
\barr
\sigma_0 &\equiv& \frac{2^9 \pi^2}{3 \exp(4)} \alpha  a_{0}^2 \approx 6.3 \times 10^{-18} \textrm{ cm}^2
\label{eq:sigma0}
\earr
is the photoionization cross-section at threshold and the function $g(\kappa)$ is such that $g(\kappa<1)=0$, $g(1) = 1$, and varies on a scale $\Delta \kappa \sim 1$ for $\kappa > 1$\cite{Bethe}. The Doppler-averaged cross-section is therefore:
\barr
\overline{\sigma}(\nu; T_{\rm m}) &=& \int_{-\infty}^{+\infty}\sigma(\nu[1 - u\Delta_{\rm H} ]) \frac{\rme^{-u^2}}{\sqrt{\pi}}\rmd u\nonumber\\
&=& \sigma_0 \int_{-\infty}^{\frac{\nu - \nuc}{\nu \Delta_{\rm H}}} g\left(\frac{\nu}{\nuc}[1 - u\Delta_{\rm H} ] \right) \frac{\rme^{-u^2}}{\sqrt{\pi}}\rmd u.
\earr
In what follows we use the distance to the Lyman limit in Doppler width units:
\beq
x \equiv \frac{\nu - \nuc}{\nuc \Delta_{\rmH}}.\label{eq:x}
\eeq
The Doppler-averaged cross-section can be rewritten as
\barr
&&\overline{\sigma}(\nu; T_{\rm m}) = \nonumber\\
&& \int_{-\infty}^{\frac{x}{1 + x \Delta_{\rm H}}} g\left([1+x \Delta_{\rm H}][1 - u \Delta_{\rm H}]\right) \frac{\rme^{-u^2}}{\sqrt{\pi}}\rmd u.
\earr
For frequencies within a few Doppler widths from the Lyman limit ($ |x| \sim $ a few), the argument of $g$ in the integral is $1 + \mathcal O(\Delta _{\rm H})$. Since $g$ varies very little on a scale $\Delta_{\rm H}$, we can set it to its threshold value in the integral, $g \approx 1$. We therefore obtain the Doppler-averaged photoionization cross section, for frequencies within a few Doppler widths of the ionization threshold:
\barr
\overline{\sigma}(\nu; T_{\rm m}) \approx \sigma_0 \left[ 1 - \frac{1}{2}\textrm{erfc}(x)\right] \equiv \sigma_0 \phi_{\rm c}(x), \label{eq:phi_c}
\earr
where the last equality defines the dimensionless profile $\phi_c(x)$.

The rate of photoionizations from the ground state per hydrogen atom per frequency interval (for the photoionizing photon) is then:
\beq
\dot{\N}_{\nu}\big{|}_{\rm phot} = - c \overline{\sigma}(\nu; T_{\rm m}) N_{\rm H}x_{1s} \N_{\nu}.
\eeq 
Detailed balance considerations show that the differential recombination coefficient to the ground state, per frequency interval for the outgoing photon (in units of cm$^3$s$^{-1}$Hz$^{-1}$) is given by:
\beq
\frac{\rmd \alpha_{1s}}{\rmd \nu} = \frac{8 \pi \nu^2}{c^2} \overline{\sigma}(\nu; T_{\rm m}) \rme^{- \frac{h(\nu - \nu_c)}{k T_{\rm m}}}\left(\frac{2 \pi \hbar^2}{m_e k T_{\rm m}}\right)^{3/2}.\label{eq:dalpha1s_dnu}
\eeq
The rate of recombinations to the ground state per hydrogen atom per frequency interval (for the emitted photon) is then
\beq
\dot{\N}_{\nu}\big{|}_{\rm rec} =  N_{\rm H} x_e^2 \frac{\rmd \alpha_{1s}}{\rmd \nu}.
\eeq
We define the continuum equilibrium occupation number:
\beq
f_{\rm eq}^c \equiv \frac{N_{\rm H} x_e^2}{x_{1s}} \left(\frac{2 \pi \hbar^2}{m_e k T_{\rm m}}\right)^{3/2}.
\eeq
The rate of change of the photon occupation number due to photoionizations and recombinations from and to the ground state, neglecting stimulated recombinations, can then be written as:
\barr
\dot{f}_{\nu}\big{|}_{\rm c} = - c \overline{\sigma}(\nu; T_{\rm m}) N_{\rm H}x_{1s} \left[f_{\nu} - \rme^{-\frac{h (\nu - \nu_c)}{k T_{\rm m}}}f_{\rm eq}^c \right].
\earr

\subsection{The radiative transfer equation in the presence of multiple overlapping lines, and photoionization and recombination from and to the ground state}
The time-dependent radiative transfer equation for the photon occupation number $f_{\nu}$, Eqs.~(\ref{eq:rad trans}), (\ref{eq:ab-em-sc alternate}), in the presence of multiple lines, is (if one line is considered as the fiducial line, then the other lines and continuum absorption and emission constitute the ``non-resonant'' processes in that equation):
\barr
&& -\frac{1}{H \nu} \frac{\partial f_{\nu}}{\partial t} + \frac{\partial f_{\nu}}{\partial \nu}= \sum_n \tau_n \varphi_n(\nu) \left[ \rme^{\frac{h(\nu - \nu_n)}{k T_{\rm m}}}f_{\nu} - f_{\rm eq}^n \right]\nonumber\\
&+& \sum_n p_{\rm sc}^n \tau_n \int \left[\varphi_n(\nu) \varphi_n(\nu') \rme^{\frac{h(\nu' - \nu_n)}{k T_{\rm m}}}- R_n(\nu, \nu')\right] f_{\nu'}\rmd \nu'\nonumber\\
&+&\frac{c \overline{\sigma}(\nu) N_{\rm H} x_{1s}}{H \nu}\left[f_{\nu} - \rme^{-\frac{h (\nu - \nu_c)}{k T_{\rm m}}}f_{\rm eq}^c\right]. \label{eq:rad trans overlap}
\earr
Since line overlap is expected to be a small correction to the recombination history, we can neglect ``corrections to the correction'' and make some approximations to simplify the calculations. As the potentially important effect is the violation of assumption (i) of the Sobolev approximation, we will lift this assumption, but keep the other three assumptions on which it relies, as we justify below.

First, as we are considering the radiation field over a characteristic frequency width a few Doppler widths, which corresponds to changes in the scale factor $\Delta a / a \sim 10^{-4}$, we can make the usual steady-state approximation by neglecting the time derivative. We also approximate the exponentials by unity since their exponents are of order $\sim \Delta_{\rm H} \frac{h \nuc}{k T_{\rm m}} \sim 10^{-3}$. In addition, we approximate the Doppler width of the Ly-$n$ line $\nu_n \Delta_{\rm H} \approx \nuc \Delta_{\rm H}$, neglecting corrections of order $\mathcal O(n^{-2})$. Finally, we will assume complete redistribution for resonant scattering. The validity of the latter approximation is more difficult to precisely quantify, but it can be justified with the following arguments. Firstly, the resonant scattering probability $p_{\rm sc}^n$ rapidly decreases as $n$ increases due to the abundance of low-energy photons that can easily photoionize atoms in the $np$ state or cause transitions to neighboring excited states. We find, for $T_{\rm m} = 3000$ K, $p_{\rm sc}^{25} = 0.44$, $p_{\rm sc}^{50} = 0.28$, $p_{\rm sc}^{100} = 0.17$ and $p_{\rm sc}^{200} = 0.09$. Secondly, for a Doppler-dominated line, partial redistribution is close to complete redistribution, in the sense that both distributions have a similar characteristic width, of the order of a Doppler width. This contrasts with the Ly$\alpha$ line, where complete redistribution can change photon frequencies by many Doppler widths due to the optical thickness of the Damping wings, and the distinction between the two types of redistribution is important.

We work with the dimensionless frequency $x$ defined in Eq.~(\ref{eq:x}). We define the optical depth for continuum absorption, per unit Doppler width:
\beq
\Phi_c(x) \equiv \tau_c \phi_c(x),
\eeq 
where
\beq
\tau_c \equiv \Delta_{\rm H}\frac{c \sigma_0 N_{\rm H} x_{1s}}{H}. \label{eq:tau_c def}
\eeq
We further define:
\barr
x_n &\equiv& \frac{\nu_n - \nuc}{\nuc \Delta_{\rm H}} = -\frac{1}{n^2 \Delta_{\rm H}} \label{eq:x_n}\\
\Phi_n(x) &\equiv& \tau_n \phi_n(x) \equiv \tau_n \frac1{\sqrt \pi}\rme^{-(x - x_n)^2}\label{eq: Phi def}\\
\Phi(x) &\equiv& \sum_n \Phi_n(x) + \Phi_{\rm c}(x)\label{eq: Phi sum}
\earr
The steady-state radiative transfer equation for $f(x)$ becomes, with the approximations justified above:
\beq
\frac{\rmd f}{\rmd x} = \sum_n \Phi_n(x) \left[f(x) - f_{\rm eq}^n \right]+\Phi_{\rm c}(x) \left[ f(x) - f_{\rm eq}^c\right] \label{eq:rad-trans-overlap}
\eeq
Since the continuum is optically thick, we set the boundary condition to $f(+\infty) = f_{\rm eq}^c$. Note that our treatment does not allow for any continuum escape in the absence of high-lying Lyman lines (the total optical depth for continuum absorption is infinite in our approximation). However, this has been shown to be negligible and lead to corrections to the ionization history $\Delta x_e/x_e \sim 10^{-6}$ \cite{CS07}.

\subsection{Generalized escape probability formalism}

\subsubsection{Preliminaries}

Let us consider the probability distribution $\Pi_{\nu}$ of photons injected with a total rate $\Gamma_{\rm inj}$ and a profile $\varphi_{\rm inj}(\nu)$ (normalized to unity), that then undergo line and continuum absorption. The evolution of the probability distribution is given by an equation similar to the radiative transfer equation Eq.~(\ref{eq:rad trans overlap}) (assuming complete redistribution and taking the exponential terms to unity):
\barr
\frac{\rmd \Pi_{\nu}}{\rmd t} &=& \frac{\partial \Pi_{\nu}}{\partial t} - H \nu \frac{\partial \Pi_{\nu}}{\partial \nu} = \Gamma_{\rm inj} \varphi_{\rm inj}(\nu) \nonumber \\
&-& H \nu \sum_n \tau_n \varphi_n(\nu) \Pi_{\nu} - c \overline{\sigma}(\nu) N_{\rm H} x_{1s}\Pi_{\nu} \label{eq:Pinu}.
\earr  
The rate at which these photons are absorbed in the Ly-$n$ transition is then $ H \nu \tau_n \int \varphi_n(\nu) \Pi_{\nu} \rmd \nu$, and the rate at which they are absorbed by the continuum is $\int c \overline{\sigma}(\nu) N_{\rm H} x_{1s}\Pi_{\nu} \rmd \nu$. Finally, these photons may also escape the set of overlapping lines by redshifting below their resonant frequencies. If we only consider lines above a given frequency $\nu_{\rm low}$ (below which lines can be considered as isolated), then the escape rate is $H \nu_{\rm low} \Pi_{\nu_{\rm low}}$. In practice, $\nu_{\rm low}$ is many Doppler widths below $\nu_c$ (so $x_{\rm low} \ll -1$), but still close enough to $\nu_c$ that we can approximate the escape rate by $H \nu_c \Pi_{\nu}(x \rightarrow -\infty)$.

Now in steady state, the sum of all these rates must equal the injection rate (there are no other possible fates than those described above for the injected photons). This can be checked explicitly by integrating Eq.~(\ref{eq:Pinu}) from $\nu_{\rm low}$ to $+\infty$ in the steady-state limit (with boundary condition $\Pi(+\infty) = 0$ since no photons are injected at infinity). Therefore, the steady-state \emph{probability} that injected photons are absorbed in the Ly-$n$ line is
\beq
P(\textrm{inj} \rightarrow n) = \frac{H \nu_c}{\Gamma_{\rm inj}} \tau_n \int \varphi_n(\nu) \Pi_{\nu} \rmd \nu,
\eeq
the probability that they cause a photoionization is 
\beq
P(\textrm{inj} \rightarrow c) = \frac{1}{\Gamma_{\rm inj}} \int c \overline{\sigma}(\nu) N_{\rm H} x_{1s}\Pi_{\nu} \rmd \nu,
\eeq
and the probability that the escape without being absorbed is
\beq
P(\textrm{inj} \rightarrow \textrm{esc}) = \frac{H \nu_c}{\Gamma_{\rm inj}} \Pi_{\nu}(x \rightarrow -\infty).
\eeq
In what follows, we apply this idea to develop a formalism for interline transition probabilities.

\subsubsection{Interline transition probabilities}

Taking advantage of the linearity of Eq.~(\ref{eq:rad-trans-overlap}), we can decompose $f(x)$ on a set of basis functions:
\beq
f(x) = \sum_j f_{\rm eq}^j \tau_j v_j(x) + f_{\rm eq}^c \int_{-\infty}^{+\infty} \Phi_c(y) G(x,y) \rmd y,
\label{eq:decomp tilde f}
\eeq
where the functions $v_j(x)$ satisfy the linear inhomogeneous differential equations:
\beq
\frac{\rmd v_{j}}{\rmd x} =  \Phi(x) v_{j}(x) - \phi_{j}(x), \label{eq:vj}
\eeq
with boundary conditions $v_j(+\infty) = 0$, and the Green's function $G(x,y)$ satisfy a similar equation:
\beq
\frac{\rmd G}{\rmd x} = \Phi(x) G(x,y) - \delta(x - y), \label{eq:Green}
\eeq
with boundary condition $G(+\infty, y) = 0$. The asymptotic behavior of $G(x,y)$ at $x \gg 1$ is:
\beq
G(x,y) \underset{x \gg 1}{\approx} \Bigg{\{}\begin{array}{ll}
0 & x>y, \\
\rme^{\tau_c(x - y)}& x<y, \end{array}   \label{eq:G infty}
\eeq
where we used $\Phi(+\infty) =  \Phi_c(+\infty) = \tau_c$. We therefore recover the appropriate boundary condition for $f(x)$ at $+\infty$:
\beq
f(x \gg 1) \approx f_{\rm eq}^c \int_x^{+\infty} \tau_c \rme^{\tau_c(x - y)} \rmd y = f_{\rm eq}^c.
\eeq
We can now use the results from the previous section. Using $\Pi_{\nu}  = v_j(x)/ (\nu_c \Delta_{\rm H})$, $ \varphi_{\rm inj}(\nu) = \phi_j(x)/(\nu_c \Delta_{\rm H})$, and $\Gamma_{\rm inj} = H/\Delta_{\rm H}$, we see that Eq.~(\ref{eq:vj}) is the steady-state version of Eq.~(\ref{eq:Pinu}). Therefore the steady-state probabilities that a photon emitted in the Ly-$j$ line is later absorbed in a Ly-$i$ transition, or subsequently photo-ionizes an atom in its ground state are, respectively: 
\barr
P(j \rightarrow i) &=& \int \Phi_i(x) v_j(x) \rmd x, \label{eq:P(i->j)}\\
P(j \rightarrow c) &=& \int \Phi_c(x) v_j(x) \rmd x.
\earr
The probability that a photon emitted in the Ly-$j$ line escapes at $x = -\infty$ without being reabsorbed in any line or causing a photoionization is given by
\beq
P(j \rightarrow \textrm{esc}) = v_j(-\infty).
\eeq
Clearly, the probability of escape from the whole set of overlapping lines is vanishingly small, except possibly for photons emitted from the lowest lying line considered as ``overlapping''. The region of line overlap blends smoothly into the region of quasi-instantaneous feedback between neighboring Lyman lines, for $n \lesssim 20-30$. Therefore, even for photons emitted from the lowest line considered as ``overlapping'' with the next higher line, there is still a near-unity probability of being reabsorbed quasi instantaneously in the next lower transition. Therefore, in practice, we have $P(j \rightarrow \textrm{esc}) = 0$ for all lines considered as overlapping, or simply close enough that feedback is quasi-instantaneous. \\
Integrating Eq.~(\ref{eq:vj}) from $-\infty$ to $+\infty$, we can see that these probabilities are complementary, as they should:
\beq
\sum_{i}P(j \rightarrow i) + P(j \rightarrow c) = 1. \label{eq:complementarity}
\eeq

The Green's function $G(x,y)$ can similarly be interpreted as the steady-state number distribution for continuum photons initially injected at the frequency $y$. The probabilities that a photon emitted at frequency $y$ is absorbed in the Ly-$i$ transition, absorbed by the continuum, or escapes at $-\infty$ are, respectively:
\barr
P(y \rightarrow i) &=& \int \Phi_i(x) G(x,y) \rmd x \label{eq:P(y->i)}\\
P(y \rightarrow c) &=& \int \Phi_c(x) G(x,y) \rmd x \label{eq:P(y->c)}\\
P(y \rightarrow \textrm{esc}) &=& G(-\infty,y).
\earr
Again, the probability for photons emitted in the continuum to escape the whole set of high-lying overlapping lines is vanishingly small, which means that in practice we have $P(y \rightarrow \textrm{esc}) = 0$.

Finally, one can check that these probabilities are indeed complementary by integrating Eq.~(\ref{eq:Green}) between $-\infty$ and $+\infty$:
\beq
\sum_i P(y\rightarrow i) + P(y \rightarrow c) = 1. \label{eq:compl.cont}
\eeq

\subsubsection{Net decay rate in the Ly-$i$ transition}
From Eq.~(\ref{eq:net decay simple}), the net decay rate in the Ly$-i$ line is
\barr
&&\dot{x}_{ip \rightarrow 1s} = 3 x_{1s} A_{ip,1s} \int \left[f_{\rm eq}^i - f(x)\right] \phi_i(x) \rmd x \nonumber\\
&&= A_{ip,1s} x_i - 3 x_{1s} \frac{A_{ip,1s}}{\tau_i} \int f(x)\Phi_i(x) \rmd x.
\earr
Using the expansion (\ref{eq:decomp tilde f}) for $f(x)$ and the definitions for the interline transition probabilities, we rewrite 
\barr
\dot{x}_{ip \rightarrow 1s}  = A_{ip,1s} x_{ip} - \frac{A_{ip,1s}}{\tau_i} \sum_j \tau_j x_{jp} P(j \rightarrow i) \nonumber\\
-3 x_{1s} \frac{A_{ip,1s}}{\tau_i}  f_{\rm eq}^c \int_{-\infty}^{+\infty} \Phi_c(y) P(y\rightarrow i) \rmd y.
\earr
We notice that $A_{ip,1s}/\tau_i \times \tau_j = A_{jp,1s}$ (approximating $\nu_i \approx \nu_j \approx \nu_c$). Also, using Eqs.~(\ref{eq:tau_n def}), (\ref{eq:dalpha1s_dnu}) and (\ref{eq:tau_c def}), we obtain that 
\beq
3 x_{1s} \frac{A_{ip,1s}}{\tau_i} f_{\rm eq}^c \Phi_c(y) = N_{\rm H} x_e^2 \frac{\rmd \alpha_{1s}}{\rmd y}, \label{eq:dalpha1s.dx}
\eeq
where $\frac{\rmd \alpha_{1s}}{\rmd x} \equiv \nu_c \Delta_{\rm H} \frac{\rmd \alpha_{1s}}{\rmd \nu}$ is the differential recombination coefficient to the ground state per unit Doppler width.\\
Using the complementarity relation (\ref{eq:complementarity}), we can now rewrite the net decay rate in the Ly-$i$ transition as:
\barr
\dot{x}_{ip\rightarrow1s} &=& x_{ip} \left(\sum_{j \neq i} R_{ip \rightarrow jp}^{(\rm ov)} + \beta_{ip}^{(\rm ov)} \right) \nonumber\\
&-& \sum_{j \neq i} x_{jp} R_{jp \rightarrow ip}^{(\rm ov)} - N_{\rm H} x_e^2 \alpha_{ip}^{(\rm ov)}, \label{eq:dot.xip.overlap}
\earr
where we have defined the overlap-induced transition rates
\beq
R_{ip \rightarrow jp}^{(\rm ov)} \equiv A_{ip,1s} P(i \rightarrow j), \label{eq:Ripjp}\\
\eeq
and the overlap-induced recombination coefficients and photoionization rates:
\barr
\alpha_{ip}^{(\rm ov)} &\equiv& \int \frac{\rmd \alpha_{1s}}{\rmd y} P(y \rightarrow i) \rmd y, \label{eq:alpha_ip.ov}\\
\beta_{ip}^{(\rm ov)} &\equiv& A_{ip,1s} P(i \rightarrow c).\label{eq:beta_ip.ov}
\earr

\subsubsection{Net rate of recombinations to the ground state}

The net rate of recombinations to the ground state, per hydrogen atom, is
\barr
&& \dot{x}_{c \rightarrow 1s} = \frac{8 \pi \nuc^2}{c^2}x_{1s} \int \overline{\sigma}(\nu)\left[\rme^{-\frac{h (\nu - \nu_c)}{k T_{\rm m}}}f_{\rm eq}^c - f_{\nu}\right] \rmd \nu \nonumber\\
&& \approx \frac{8 \pi \nuc^2}{c^2}x_{1s} \nuc \Delta_{\rm H} \sigma_0 \int \phi_c(x)\left[f_{\rm eq}^c - f(x)\right] \rmd x,\label{eq:cont-1s}
\earr
where in the second line we took $\rme^{-\frac{h(\nu - \nuc)}{k T_{\rm m}}} \approx 1$ in the vicinity of the Lyman limit. Using again the decomposition (\ref{eq:decomp tilde f}) and the definitions of the interline transition probabilities, we get:
\barr
\dot{x}_{c \rightarrow 1s} &=& \frac{8 \pi \nuc^2}{c^2}x_{1s} \nuc \Delta_{\rm H} \sigma_0 f_{\rm eq}^c \times \Gamma_c \nonumber\\
&-&\frac{8 \pi \nuc^3}{3 c^2}\Delta_{\rm H} \sigma_0 \sum_j x_{jp} \frac{\tau_j}{\tau_c}P(j\rightarrow c),
\earr
where we have defined
\barr
\Gamma_c &\equiv& \int \phi_c(x) \left[ 1 - \int \Phi_c(y) G(x,y) \rmd y \right] \rmd x \nonumber\\
&=& \int \rmd x~\phi_c(x) \int \rmd y \left[ \delta(y - x) - \int \Phi_c(y) G(x,y) \right] \nonumber\\
&=& \int \rmd y \int \rmd x ~ \phi_c(x) \left[ \delta(y - x) - \int \Phi_c(y) G(x,y) \right] \nonumber \\
&=& \int \phi_c(y) \left[1 - P(y \rightarrow c) \right] \rmd y\nonumber\\
&=& \int \phi_c(y) \sum_j P(y \rightarrow j), 
\earr
where we have used the definition (\ref{eq:P(y->c)}) of $P(y\rightarrow c)$ in the fourth line, and in the last line we have used the complementarity relation (\ref{eq:compl.cont}).

After some algebraic manipulations, we can cast the net rate of recombinations to the ground state in the following form:
\beq
\dot{x}_{c \rightarrow 1s} = \sum_j \left[N_{\rm H} x_e^2 \alpha_{jp}^{(\rm ov)} - x_{jp} \beta_{jp}^{(\rm ov)}\right],\label{eq:dot.xc.overlap}
\eeq
where the overlap-induced recombination coefficients and photoionization rates have been defined in Eqs.~(\ref{eq:alpha_ip.ov}) and (\ref{eq:beta_ip.ov}).

\subsubsection{Rate of change of the ground state population}\label{sec:dot.x1s.overlap}
Adding Eqs.~(\ref{eq:dot.xip.overlap}) and (\ref{eq:dot.xc.overlap}), we see that the net rate of change for the ground state population due to decays from the overlapping lines and recombination to the ground state is zero: 
\beq
\dot{x}_{1s}^{(\rm ov)} = - \dot{x}_e^{(\rm ov)} = \sum_{j,~\rm ov} \dot{x}_{jp\rightarrow 1s} + \dot{x}_{c \rightarrow 1s} = 0.
\eeq
This can be understood intuitively since any photon emitted from one of the overlapping lines is bound to be reabsorbed in a neighboring line or photoionize an atom from the ground state; similarly, a photon emitted after a recombination to the ground state will almost certainly be reabsorbed in a high-lying Lyman line, or cause a subsequent photoionization. In other words, no escape is possible from the series of high-lying lines, and the net decay rate to the ground state therefore vanishes. Line overlap can therefore only influence the recombination history indirectly, through changing the populations of the excited states.

We now turn to the numerical evaluation of the overlap-induced transition rates.

\subsection{Evaluation of  the overlap-induced transition rates} \label{section:overlap evaluation}

\begin{figure} 
\includegraphics[width = 85mm]{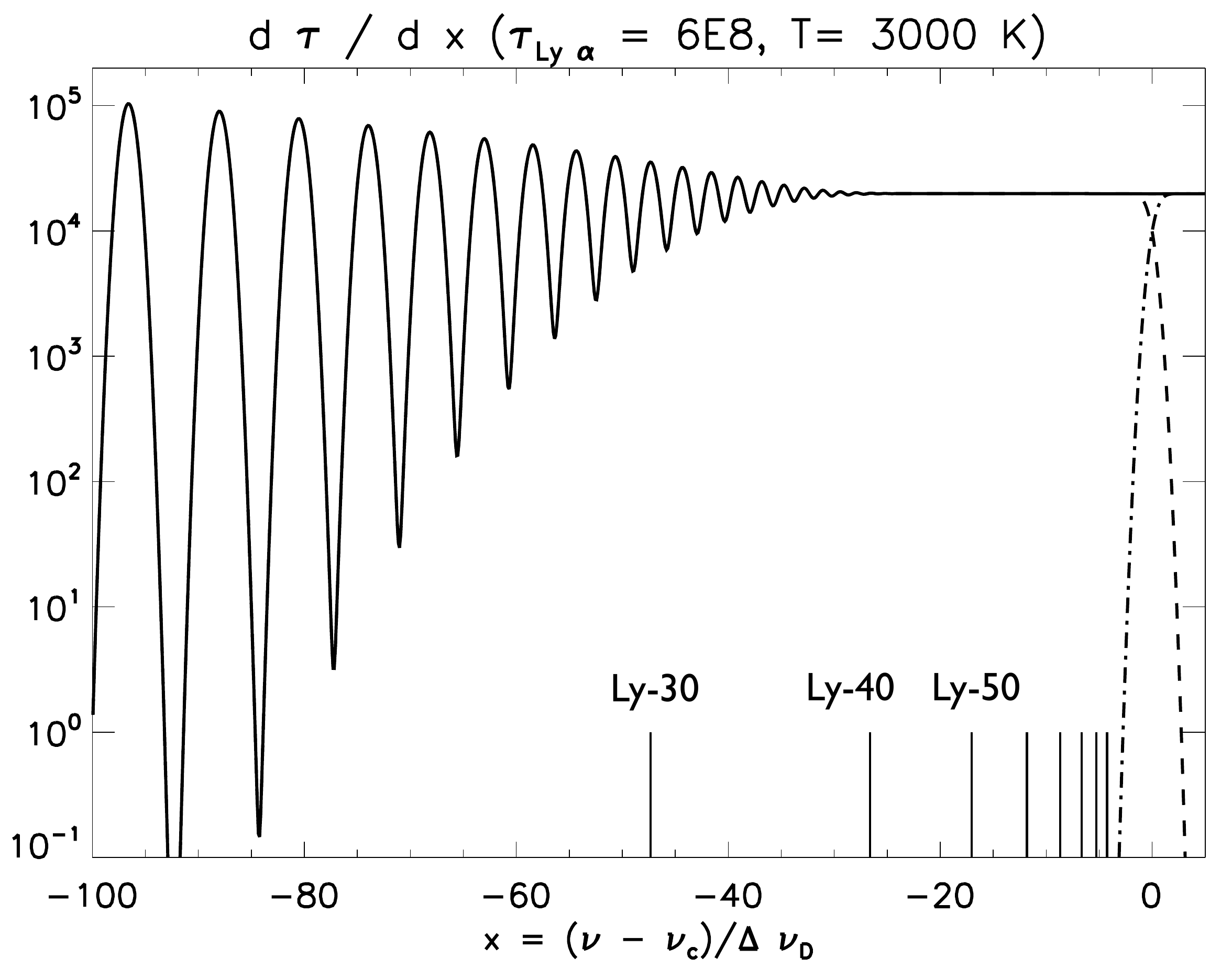}
\caption{Differential optical depth per unit Doppler width, $\Phi(x)$, for fiducial values $\tau_{\Ly \alpha} = 6\times 10^8$ and $T_{\rm m} = 3000$ K. The dashed line is the contribution from Lyman lines, the dot-dashed line is the contribution from photoionization from the ground state, and the solid line is their sum. The position of the Lyman-30 to 100 resonant frequencies (spaced by $\Delta n = 10$) is shown.}
\label{fig:Phi_x}
\end{figure}

We show in Fig.~\ref{fig:Phi_x} the optical depth per unit Doppler width, $\Phi(x)$. It can be seen that for $x \gtrsim x_{n_{\rm ov}}$, $\Phi(x)$ is nearly constant. This can be explicitly derived as follows. From Eqs. (\ref{eq:Anp1s}), (\ref{eq:tau_n def}), (\ref{eq:sigma0}) and (\ref{eq:tau_c def}), we can show that
\beq
\tau_n \underset{n \gg 1}{\sim}  \frac{2}{\Delta_{\rm H} n^3} \tau_c.
\eeq
Moreover, the separation between two neighboring high-lying lines has the asymptotic expression
\beq
x_{n+1} - x_n \sim \frac{2}{\Delta_{\rm H} n^3}.
\eeq
For $x \gtrsim x_{n_{\rm ov}}$ (such that the separation between neighboring lines becomes small compared to unity), we can approximate the sum of optical depths due to Lyman transitions as a Riemann integral:
\barr
\sum_n \tau_n \phi_n(x) &\approx& \tau_c \sum_n \left(x_{n+1} - x_n\right) \frac{1}{\sqrt \pi} \rme^{-(x - x_n)^2} \nonumber\\
 &\approx& \tau_c \frac1{\sqrt \pi}\int_{-\infty}^0 \rme^{-(x - u)^2} \rmd u \nonumber\\
&=& \frac{\tau_c}{2} \textrm{erfc}(x) = \tau_c - \tau_c \phi_c(x).
\earr
Thus, we obtain, using the definition (\ref{eq: Phi sum}):
\beq
\Phi(x \gtrsim x_{n_{\rm ov}}) \approx \tau_c.
\eeq
Equation (\ref{eq:vj}) therefore has an analytic solution:
\barr
v_j(x) &=& \frac{1}{2} \rme^{-(x - x_j)^2}\textrm{erfc}\left(x - x_j +\frac{\tau_c}{2}\right)\rme^{\left(x - x_j + \frac{\tau_c}{2}\right)^2}\nonumber\\
&\approx& \frac{1}{\tau_c} \frac{\rme^{-(x - x_j)^2}}{\sqrt{\pi}}
\label{eq:uj approx large j}
\earr
where in the second line we used $\tau_c \gg 1$ ($\tau_c \gtrsim 10^2$ at all times). \\
We therefore obtain:
\barr
P(j \rightarrow i) &\approx& \frac{\tau_i}{\tau_c} \frac{1}{\sqrt{2 \pi}} \rme^{-\frac{1}{2}(x_i - x_j)^2} \nonumber\\
 &\approx& \frac{2}{\Delta_{\rm H} i^3}\frac{1}{\sqrt{2 \pi}} \rme^{-\frac{1}{2}(x_i - x_j)^2} \label{eq:Pj->i}\\
P(j \rightarrow c) &\approx& 1 - \frac{1}{2 \sqrt \pi}\int_{-\infty}^{+\infty}\textrm{erfc}(x) \rme^{-(x-x_i)^2} \rmd x \nonumber\\
&=& 1 - \frac12 \textrm{erfc}\left(\frac{x_j}{\sqrt 2}\right),
\earr
where the integral can be evaluated after differentiating with respect to $x_i$. The overlap-induced transition rates $R_{ip\rightarrow jp}^{(\rm ov)}$ and photoionization rates $\beta_{jp}^{(\rm ov)}$ can then be obtained from Eqs.~(\ref{eq:Ripjp}) and (\ref{eq:beta_ip.ov}).

Similarly, the function $v_c(x) \equiv \int \phi_c(y) G(x,y) \rmd y$ satisfies the following differential equation:
\beq
\frac{\rmd v_c}{\rmd x} = \Phi(x) v_c(x) - \phi_c(x),
\eeq
which has the solution, valid to lowest order in $1/\tau_c$:
\beq
v_c(x) \approx \frac{1}{\tau_c}\left[1 - \frac12 \textrm{erfc}(x)\right].
\label{eq:uc approx}
\eeq
The overlap-induced recombination coefficient can then be written as:
\beq
\alpha_{ip}^{(\rm ov)} = \frac{8 \pi \nu_c^3 \Delta_{\rm H}}{c^2} \sigma_0 \left(\frac{2 \pi \hbar^2}{m_e k T_{\rm m}}\right)^{3/2}\int \Phi_i(x) v_c(x) \rmd x,  
\eeq
where we used Eqs.~(\ref{eq:P(y->i)}), (\ref{eq:dalpha1s.dx}) and (\ref{eq:alpha_ip.ov}).

After some manipulations, we can show that the overlap-induced recombination coefficients can be simply expressed in terms of the overlap-induced photoionization rates: 
\beq
\alpha_{ip}^{(\rm ov)} \approx 3 \left(\frac{2 \pi \hbar^2}{m_e k T_{\rm m}}\right)^{3/2} \beta_{ip}. \label{eq:overlap.alpha.db}
\eeq
From the asymptotic expression of $A_{jp,1s} \propto j^{-3}$, we can also show that
\beq
R_{jp\rightarrow ip} \approx R_{ip \rightarrow jp}.\label{eq:overlap.R.db}
\eeq
Equations (\ref{eq:overlap.alpha.db}) and (\ref{eq:overlap.R.db}) are simply the usual detailed balance relations, in the limit $\nu_i \approx \nu_j \approx \nu_c$.

The expressions provided in this section are valid for $n \gtrsim n_{\rm ov}$, when Lyman lines are within less than a Doppler width of each other. We therefore expect the expressions for $\beta_{np}^{(\rm ov)}$ and $\alpha_{np}^{(\rm ov)}$ to be accurate in the regime where they are significant, for $n \gtrsim n_{\rm ov,c} \gg n_{\rm ov}$. On the other hand, the interline transition probabilities $P(i\rightarrow j)$ should smoothly transition from the asymptotic expression (\ref{eq:Pj->i}) for $i,j \gtrsim n_{\rm ov}$ to the Sobolev values for nearly instantaneous feedback for $i, j \lesssim n_{\rm ov}$, that is 
\beq
P(i \rightarrow j) = \tau_i^{-1} \delta_{j,i-1} +  \left(1 - \tau_i^{-1}\right) \delta_{ij} \approx \delta_{ij}.
\eeq
We checked that this is indeed the case by integrating numerically Eqs.~(\ref{eq:vj}) and (\ref{eq:P(i->j)}). We therefore set $P(i\rightarrow j) = \delta_{i,j}$ for $\min(i,j) < n_{\rm ov}$, and use Eq.~(\ref{eq:Pj->i}) otherwise. Since, as we shall see below, line overlap appears to lead to negligible changes to the recombination history, the exact value of the interline transition probabilities near $n_{\rm ov}$ is not critical.

\subsection{Results}

As we showed in Section \ref{sec:dot.x1s.overlap}, the net rate of change of the ground state population through the high-lying Lyman transitions vanishes. Since any $np \rightarrow 1s$ transition is systematically followed by the absorption of the emitted photon, the high-lying $np$ states are virtually radiatively connected to one another (and to the continuum), rather than being radiatively connected to the ground state. In the language of Ref.~\cite{EMLA}, the high-lying $p$ states are \emph{interior} states. The EMLA formalism developed in Ref.~\cite{EMLA} can then easily be extended to include the overlap-induced transitions, which only depend on the matter temperature.

We added the overlap-induced H$(np) \leftrightarrow \textrm{H}(n'p)$ and H$(np) \leftrightarrow e^-+p$ transitions to our effective rates code. We computed the change in the effective recombination coefficients $\mathcal{A}_{2s}$, $\mathcal{A}_{2p}$ and effective $2p \rightarrow 2s$ transition rate $\mathcal{R}_{2p \rightarrow 2s}$ when the states $2s,~ 2p$ are considered as the only interface states (i.e. cutting off the Ly$\beta$ transition and above for simplicity; the effect of overlap is independent of that simplification). We find that the relative changes in each of the effective coefficients are at most a few times $10^{-5}$. As a comparison, the change in total effective recombination coefficient between $n_{\max} = 100$ and $n_{\max} = 200$ is of order $0.3$ to $2\%$ over the temperature range considered; from $n_{\max} = 200$ to $n_{\max} = 400$, this change is of order $0.08$ to $ 0.6\%$. Therefore, the effect of overlap is a few orders of magnitude smaller than the mere error due to the necessary truncation of the high energy shells when computing the effective rates. Since previous work \cite{Grin_Hirata, Chluba_Vasil} have shown that MLA computations with $n_{\max} \sim 100$ already reach the desired level of accuracy, we conclude that line overlap can be safely ignored. 


\section{Population Inversion}
\label{minimestepawayfromthelaser}
Another important effect to consider 
is that of population inversion.
Sufficiently bottlenecked multi-level systems may develop population inversion between radiatively coupled states. If this effect is dramatic, if the velocity field is sufficiently coherent, and the effective path length long enough, stimulated emission may lead to intense, narrow, coherent \textit{maser} radiation \cite{elitzur,lo}. Many astrophysical masers are now known, and play an important role in firming up the extra-galactic distance scale \cite{pogreb,comet,lo,binneym,elitzur,strel2,freedman}. Most known astrophysical masers are molecular, but the emission-line star MWC349 is one example of a hydrogen recombination line maser \cite{strel1,strel2}. 

It has been suggested both that the recombining primordial hydrogen plasma exhibits sufficient population inversion for an all-sky natural maser at high $n$, and that rare, extremely overdense regions mase during the epoch of recombination \cite{spnorman,scott_laser}. The possibility of a cosmic recombination maser is particularly enticing, because it could conceivably amplify weak low-frequency $\nu=100~{\rm Mhz}-1~{\rm Ghz}$ CMB spectral distortions from recombination to a detectable level \cite{RMCS06}. 

Output from \textsc{RecSparse} computations shows that some radiatively connected $\alpha$-transitions ($\Delta n=\pm 1$) between states with $l=\mathcal{O}\left(1\right)$ and $n\sim50$ do show rather dramatic population inversion from $z\lesssim 800$ onward. We have verified that the width of lines showing population inversion is dominated by Doppler and not natural broadening. If we neglect 
natural broadening, all pairs of radiatively connected $\Delta l=\pm 1$ pairs for some fixed values of $n,n^{\prime}$ (where $n$ and $n^{\prime}$ denote the hydrogen shells connected by the line) have the same line profile, $\phi_{n,n^{\prime}}\left(x\right)$. In this case, the steady-state radiative transfer equation may be solved to obtain the photon occupation number
$f^{-}_{n,n^{\prime}}$ on the red side of the line in terms of the occupation number $f^{+}_{n,n^{\prime}}$ on the blue side of the line\footnote{The method is analogous to that used to obtain Eqs.~(\ref{eq:rad transfer sobolev})-(\ref{sobfinal}), with the modification of injection by multiple transitions with the same $n$ and $n^{\prime}$. The result is derived in detail in Ref. \cite{grinthesis}.}:
\begin{align}
&f^{-}_{n,n^{\prime}}=f^{+}_{n,n^{\prime}}+\left(f^{{\rm eq}}_{n,n^{\prime}}-f^{+}_{n,n^{\prime}}\right)\left(1-e^{-\tau_{n,n^{\prime}}}\right),\label{laser_anal_result}\\
&\tau_{n,n^{\prime}}\equiv \sum_{l,l^{\prime}}\tau_{n,n^{\prime}}^{l,l^{\prime}},\\
&f^{\rm eq}_{n,n^{\prime}}\equiv \frac{\sum_{l,l^{\prime}}\tau_{n,n^{\prime}}^{l,l^{\prime}}f^{{\rm eq},l,l^{\prime}}_{n,n^{\prime}}}{\sum_{l,l^{\prime}}\tau_{n,n^{\prime}}^{l,l^{\prime}}},
\end{align}
where $\tau_{n,n^{\prime}}^{l,l^{\prime}}$ is the optical depth in the transition whose initial and final states have quantum numbers $(n,l)$, and $(n^{\prime},l^{\prime})$. We use the convention $n^{\prime}>n$. The sum is over all allowed sets of quantum numbers obeying dipole selection rules. The occupation number in equilibrium with this transition is $f^{{\rm eq},l,l^{\prime}}_{n,n^{\prime}}\equiv x_{n^{\prime},l^{\prime}}g_{l}/\left(x_{n,l}g_{l^{\prime}}\right)$, and $g_{l}=2(2l+1)$ is the statistical degeneracy of an atomic state with angular momentum $l$.

It is easily seen from Eq.~(\ref{laser_anal_result}) that significant amplification due to stimulated emission requires $\tau_{n,n^{\prime}}\ll -1$. We inspect the output of a \textsc{RecSparse} run with $n_{\rm max}=180$ and find that when $\tau<0$, $|\tau|< 10^{-8}$ in the \textit{most} amplified case.  In the purely radiative case, then, there is no high-$n$ cosmological hydrogen recombination maser.\\

The populations  of the $2p$ and $2s$ states are strongly out of equilibrium towards the end of the epoch of primordial recombination, with the ratio $x_{2p}/3x_{2s}$ reaching values as large as $\sim 3$ for z $\sim 500-600$. Although we do not track the populations of the $2p_{3/2}$ and $2p_{1/2}$ states separately, the overlap of the $1s_{1/2}- 2p_{1/2}$ and $1s_{1/2}- 2p_{3/2}$ Ly$\alpha$ doublet ensures that they are in statistical equilibrium, so $x_{2p_{3/2}}/2 x_{2s_{1/2}} = x_{2p}/3x_{2s}$. It is therefore \emph{a priori} possible for the 11GHz $2p_{3/2} \rightarrow 2s_{1/2}$ transition to be amplified. We checked however that the largest negative optical depth in this transition is $\tau \sim -10^{-7}$ for $z \sim 1100$, and therefore the population inversion does not lead to any significant amplification.

\section{Conclusions}\label{section:discussion}
In this work, we have evaluated the impact of several previously neglected radiative transfer effects on cosmological hydrogen recombination:

$\bullet$ \emph{Thomson scattering in the Ly-$\alpha$ line} was shown to be negligible, with corrections to the recombination history $|\Delta x_{e}/x_{e}| \lesssim 3 \times 10^{-5}$. We showed that at early times, $z \gtrsim 1300$, the dominant effect was a \emph{delay} of recombination, due to the reinjection of photons from the red side of the line back into the blue side of the line during large angle scattering events. That effect can only be properly accounted for with a full kernel approach, since scattered photons are redistributed on frequency scales larger than the characteristic scale over which the radiation field changes. At lower redshifts, recoil becomes dominant and Thomson scattering accelerates recombination by helping photons escape from the Ly$\alpha$ line. 

$\bullet$ \emph{Distortions from the deuterium Ly-$\alpha$ line} were shown to be negligible. Indeed, the very fast D$(2p) \rightarrow$ D$(1s)$ transition rate, due to the relatively small optical depth in the deuterium line, brings the deuterium $2p$ to $1s$ ratio close to equilibrium with the incoming radiation field on the D Ly$\alpha$ line. Moreover, any distortions are further washed out due to the very large frequency diffusion rate caused by resonant scattering by neutral hydrogen. Accounting for deuterium therefore leads to changes of at most $\mathcal{O}(10^{-5})$ to the recombination history due to the small change in the expansion history and the ambiguity in defining $x_e$.

$\bullet$ \emph{The high-lying, non-overlapping Lyman transitions} above Ly$\gamma$ (strictly) can be artificially cut off without loss of accuracy. Only the $2s, 2p, 3p$ and $4p$ states therefore need to be considered as ``interface'' states in an EMLA computation \cite{EMLA}. Diffusion in Ly$\beta$ and higher lines can be neglected.

$\bullet$ \emph{Overlap of the high-lying Lyman lines}, as well as overlap of the extremely high-lying lines with the continuum, was shown to lead to $\mathcal{O}(10^{-5})$ changes to the effective transition rates. This change is a few orders of magnitude smaller than the mere truncation error in the effective transition rates computed with $n_{\max} \sim 100$ energy shells. The effect of line overlap is therefore negligible.

$\bullet$ \emph{Cosmological hydrogen masers} are shown not to arise in this purely radiative treatment.

The goal of the ongoing work in the field is to develop a complete theory for hydrogen recombination, with a well understood error budget. In this paper, we have evaluated the impact of some radiative transfer effects that had not been previously addressed. While it is possible that some effects have not been considered yet, we believe that most of the radiative transfer effects relevant in primordial hydrogen recombination are now well understood. The picture is less clear for the effect of collisional processes \cite{Chluba_Vasil}, for which the rates are relatively poorly known. Ultimately, we need a recombination code that is not only accurate, but also fast, in order to be included in Markov chains for cosmological parameter estimation; this will be the subject of future work.

\section*{Acknowledgments}
The authors thank Jens Chluba for useful discussions about Thomson scattering in Lyman-$\alpha$ and aknowledge fruitful conversations with the participants of the July 2009 Paris Workshop on Cosmological Recombination. Y. A-H. and C. H. are supported by the U.S. Department of Energy (DE-FG03-92-ER40701) and the National Science Foundation (AST-0807337). D.G. is supported by the Dan David Foundation, the Gordon and Betty Moore Foundation, and the National Science Foundation (AST-0807044). C. H. is supported by the Alfred P. Sloan Foundation.

\appendix

\bibliography{references}

\end{document}